\documentclass[11pt]{article}
\usepackage{amssymb,amsmath,graphicx}   

\def\mb#1{{\mbox{\boldmath $#1$}}}

\newcommand{\be}{\begin{eqnarray}}
\newcommand{\nd}{\end{eqnarray}}

\begin{document}
\noindent{\Large{\bf A REVIEW ON FISH SWIMMING AND BIRD/INSECT FLIGHT}}
\vskip 2mm
\noindent{Theodore Yaotsu Wu\\
{\it Engineering and Applied Science, California Institute of Technology\\
Pasadena, California 91125; email: tywu@caltech.edu}
\vskip 2mm
\noindent{\bf Key Words} ~~~fish swimming, bird/insect flight, reactive theory, bound vortex shedding, free wake vortices,  power, thrust, energy conservation, unsteady slender-body theory, acceleration potential, fully nonlinear unsteady flexible wing theory, insect wing leading-edge-vortex high lift, locomotion energetics, metabolic rates, scale effects
\vskip 2mm
\noindent$\bullet$ {\bf Abstract}  ~This expository review is devoted to fish swimming and bird/insect flight.  (i) The simple waving motion of an elongated flexible ribbon plate of constant width propagating a wave distally down the plate to swim forward in a fluid, initially at rest, is first considered to provide a fundamental concept on energy conservation.  It is generalized to include variations in body width and thickness, with appended dorsal, ventral and caudal fins shedding vortices to closely simulate fish swimming, for which a nonlinear theory is presented for large-amplitude propulsion.  (ii) For bird flight, the pioneering studies on oscillating rigid wings are discussed with delineating a fully nonlinear unsteady theory for a 2D flexible wing with arbitrary variations in shape and trajectory to provide a comparative study with experiments.  (iii) For insect flight, recent advances are reviewed by items on aerodynamic theory and modeling, computational methods, and experiments, for forward and hovering flights with producing leading-edge vortex to yield unsteady high lift. (iv) Prospects are explored on extracting prevailing intrinsic flow energy by fish and bird to enhance thrust for propulsion.  (v) The mechanical and biological principles are drawn together for unified studies on the energetics in deriving metabolic power for animal locomotion, leading to the surprising discovery that the hydrodynamic viscous drag acting on swimming fish is largely associated with laminar boundary layers, thus drawing valid and sound evidences for a resounding resolution to the long-standing fish-swim paradox proclaimed by Gray (1936, 1968). 

\vskip 1.8mm
\noindent{\bf 1. INTRODUCTION}
\vskip 1mm
Fish swimming and bird flight are of agelong human interests, with their fascinating nature to appreciate, diversity to behold, and complexities to comprehend.  The focal inspiration for the interest is curiosity.  That is the seed of creativity, thereby bringing forth fruitful formation of sound physical conceptions,  successful developments of required mathematical tools, and novel engineering.  Without curiosity, one may wonder what would have become to riding in jet airplanes and in cruising ships across oceans for the ease and comfort in transportation.  This is how the advances in science and technology have been all about.
\vskip 1mm
Oscillatory airfoil theory has been developed since 1920s by such pioneers as Wagner (1925), Theodorsen (1935), K$\ddot{u}$ssner (1935) and others, given in one or another form of linear theory.  The application was first to aero-elasticity problems to expound the fluttering forces and their effects on airplane wing-fuselage system flying through turbulent and gusty air.  The fundamental elements of the theory would also be applicable to studies on fish swimming and bird flight.
\vskip 1mm
Keen interests have been stimulated by the paper published by G.I. Taylor (1952) on the swimming of long and narrow animals, arousing research activities at various institutions, especially at Cambridge University and Caltech.  Close academic interactions and collaborations between various centers plus their prolific publications rendered it timely to hold the first ``International Symposium on Swimming and Flying in Nature" in Aug. 1974 at Caltech, participated by $250$ of the active workers in the field.  In Sept. 1975, another symposium, called by Lighthill and Weis-Fogh, was held at Cambridge University on ``Scale Effects in Animal Locomotion", a new subject of great significance.  These symposia are of value to have set fruitful directions for vital advances in this growing important field.
\vskip 1mm
For fish propulsion, the first conspicuous feature is the undulatory motion of the flexible fish body passing a wave distally in swimming.  This feature is abstracted to expound the simplified geometry of an elongated flexible ribbon plate of constant width, to which the slender-body analysis is applicable as presented in \S 2.1-2.3.  The fundamental quantities derived stepwise in this simple case provide a crisp clear physical concept on momentum, power expended for generating thrust on the plate with some energy expelled at an optimum efficiency for the plate to swim forward.  These primary results are then generalized in \S 2.4.1-2.4.3, by including all the other main factors to simulate fish swimming in nature on general slender-body theory given in \S 2.5.  To this end, the nonlinear theory introduced by Lighthill (1971) for large-amplitude swimming is finally developed in this work to completion in \S 2.6.
\vskip 1mm
For bird flight, the classical oscillating airfoil theory is discussed in \S 3.1, followed by presenting a fully nonlinear unsteady theory developed by Wu (2001b)-(2007) for a flexible two-dimensional wing with arbitrary variations in wing shape and its trajectory as summarized in \S 3.2, thus furnishing a comparative study by comparisons with two well noted experiments on heaving and pitching of airfoils, respectively.  The insect flight is reviewed in \S 4, itemized for the aerodynamic theory and modeling in \S 4.1, computational fluid dynamics in \S 4.2, and experimental investigations in \S 4.3.
\vskip 1mm
For fish swimming and bird flight in wild nature, e.g. in torrential rivers and in whirling winds, the prospects are explored in \S 5 on the animal's apt capabilities of extracting intrinsic flow energy to gain more thrust for propulsion in unfavorable environments with aspects relevant to control theory.
\vskip 1mm
For the energetics of animal locomotion examined in \S 6, the most effective and efficient parts of mechanics (on reactive theory for determining thrust) and biology (on measuring the metabolic rate of converting oxygen consumption into muscular power) are drawn together for unified studies on the total energy conservation and the balance between the thrust exerted by the fish and the hydrodynamic viscous drag acting on the fish.  The highly rewarding results are as just summed in Abstract.

\vskip 3.8mm
\noindent{\bf 2. FISH SWIMMING}
\vskip 1mm
In general, a fish swimming in water adopts a sidewise undulatory waving motion of its flexible body of length $\ell$ and depth $2b$ say, commonly appearing in the form of the so-called slender body, or elongated body, typified  by the slenderness-ratio $\delta=b/\ell\ll 1$.  The waving motion in general propagates a wave distally down the body to generate a forward thrust $T$ to balance out the hydrodynamic viscous drag $D$ acting on the fish so as to sustain a steady swimming speed $U$ as so resulted.  For most differing species of grown-up fishes in cruising swim, the Reynolds number $Re=U\ell/\nu ~(\nu$ being the kinematic viscosity of the fluid) would be so high ($\simeq 10^3$ for small slow fish, $\simeq 10^8$ for whales) and the boundary layer enveloping the fish so thin that its thickness can be neglected for evaluation of thrust $T$ and power $P$ of fish locomotion, a powerful method known as the {\it reactive theory of animal locomotion}, which is commonly adopted with success.  With body motion specified and analyzed with certain viscous drag distributed along the body, it may lead to found a {\it resistive theory of animal locomotion}, of which the basic ideas have been delineated by Lighthill (1973a, \S 2).  On the other hand, the bio-chemical studies of the oxygen consumption by a steadily swimming fish and its metabolic conversion into muscular energy for swimming is a science and technology well achieved, as shown in their precious contributions to this problem.  Perhaps due to this contrast in recognizing their early status quo, a highly inspiring {\it paradox on fish locomotion} was proclaimed by Sir James Gray (1936, 1968).  Briefly, it asserts that {\it \underline{either} fish can experience a viscous drag much smaller than the drag acting on their artificial smooth model \underline{or} they can deliver muscular power in sustained swimming an order of magnitude greater than that delivered by the same mass of muscle of warm-blooded animals working at the same level of activity}.  This claim has provided a strong stimulus to have attracted both mechanical and biological studies on fish locomotion.
\vskip 2mm
\noindent{\bf 2.1. Simplified Reactive Swimming}
\vskip 1mm
In aquatic animal locomotion, the first conspicuous feature to be noted is the flexibility of fish body.  An early study on this feature is a Fourier synthesis of a waving plate pursued by Wu (1961), with plate undulation represented by a Fourier series in general.  In addition, another feature of acceleration of waving plate was expounded by Wu (1962).  As explained, these general geometric shape functions are first abstracted here to a generic undulatory plate of a very simple geometry for fundamental study.  Thus, we consider the propulsion, on reactive theory, of a flexible ribbon plate of constant width spanning $-b\leq y\leq b$, initially stretched straight from the leading edge at $x=0$ to the trailing edge at $x=\ell~(\delta=b/\ell\ll 1)$, with its thickness neglected.  It then starts from time $t=0$ to propagate a continuous waving motion given in cartesian coordinates $\mb{x}=(x,y,z)$ fixed in the body frame as
\be  z=h(x,t) \qquad~(0<x<\ell,~|y|< b,~t>0), \label{1}\nd
with the plate immersed in a fluid of density $\rho$, which is unbounded and has a uniform free stream velocity $(U, 0, 0)$ far away.  Here, $U$ is the swimming velocity of the plate in the $-x$-direction in the absolute frame with fluid at rest far away.  The fluid is assumed incompressible and inviscid, and the resulting flow assumed irrotational except for the vortex sheet generated at the ribbon surface and the vortex sheet shed from the trailing edge to become free vortices drifting away with the local fluid.
\vskip 1mm
For flow analysis, we adopt the `{\it linear slender-body theory}' applicable to bodies of $\delta=b/\ell\ll 1$ moving with gradual gradients so that the fluid motion at arbitrary station $x$ is all determined in the $yz$-plane transverse to the $x$-direction.  Thus, the flow at a given $x$ at time $t$ depends only on its own $h(x,t)$ and otherwise unaffected by the plate motion both up- and down-stream.
\vskip 1mm
Thus, originally, the 3D flow has a perturbation velocity potential $\phi (x,y,z)$ such that
\begin{subequations}\label{2}
\be  \nabla\phi =(u,v,w), \quad~\mb{q}=(U+u,v,w), \quad~ \nabla^2\phi=0, ~~\mbox{and}~~ \phi\rightarrow 0~~\mbox{as}~~|\mb{x}|\rightarrow \infty,  \label{2a}\nd
where $\mb{q}$ is the total velocity vector, and the Bernoulli equation for pressure $p$ is given by
\be  p/\rho =-D\phi-\frac{1}{2}(\nabla\phi)^2, \quad~~ D=\frac{\partial}{\partial t}+U\frac{\partial}{\partial x}. \label{2b}\nd\end{subequations}
\vskip 1mm
However, for the linearized slender-body flow in the $yz$-plane, or in the complex $\zeta=(y+iz)$-plane, there exist a complex velocity potential $f=\phi +i\psi$ and complex velocity $\chi=df/d\zeta=v-iw$, or
\be \zeta=y+iz,~~~f=\phi +i\psi,~~~\chi=df/d\zeta=v-iw; \quad~ v_y+w_z=0, ~~~v_z-w_y=0, \label{3}\nd
where $\zeta,~f$ and $\chi$ are analytical functions of one another under the incompressibility and irrotationality conditions as expressed by the final two relations in (3), where $v_y=\partial_y v=\partial v/\partial y$, etc. 
\vskip 1mm
The plate motion ({1}) induces a flow velocity field with its $z$-component at the plate given by
\be w_{\pm}=w(x,y,h\pm 0,t)=Dh(x,t)=(\partial_t+U\partial_x)h \equiv W(x,t) \quad~~(|y|<b). \label{4}\nd
With this boundary condition at the plate and with $\chi\rightarrow 0$ as $|\zeta|\rightarrow\infty$, the flow has the solution:
\be  \chi=v-iw=iW\left(\frac{\zeta}{\sqrt{\zeta^2-b^2}}-1\right),\quad f=\phi +i\psi= iW(\sqrt{\zeta^2-b^2}-\zeta)~~~(\phi_\pm=\mp W\sqrt{b^2-y^2}).\label{5}\nd
The flow momentum has zero $y$-component since $v$ is odd in $y$, and its $z$-component per unit longitudinal length (or unit thickness in $x$) is given by
\be {\cal M}=\rho\int\int_S \frac{\partial\phi}{\partial z} dydz=\rho\int (\phi_- -\phi_+)dy =2\rho W\int_{-b}^b \sqrt{b^2-y^2} dy=mW(x,t) \quad~(m=\rho\pi b^2), \label{6}\nd
where the $S$-integral is over the entire $yz$-plane per unit length in $x$, giving, by the $z$-integral, the jump $(\phi_- -\phi_+)$ which exists only over the plate, hence the result, with $m=\rho\pi b^2$ signifying the added mass (or the virtual mass) of the fluid at station $x$.  In terms of flow singularities, this momentum is due to an elliptic distribution of doublets (or dipoles) at the plate, namely $(\phi_--\phi_+)=2W\sqrt{b^2-y^2}$.
\vskip 1mm
The instantaneous lift per unit length of the plate is the force, positive in the $z$-direction, called the {\it specific lift}, ${\cal L}(x,t)$, given by
\be  {\cal{L}}(x,t)=\int_{-b}^b (p_--p_+)dy=-\rho\int_{-b}^b D(\phi_- -\phi_+)dy = -D(mW)=-D{\cal M}. \label{7}\nd
Then the instantaneous power $P$ exerted by the plate in self-propulsion, the kinetic energy $E$ of the fluid, and the thrust $T$ being the $-x$-component of the lift acting over the waving plate are given by
\be P=-\int_0^\ell \frac{\partial h}{\partial t}{\cal{L}}dx, \quad~ E=-\int_0^\ell \left(\frac{\partial h}{\partial t}+U \frac{\partial h}{\partial x}\right){\cal{L}}dx, \quad~ T=\int_0^\ell \left(\frac{\partial h}{\partial x}\right) {\cal L}dx, \label{8}\nd
which signify that $P$ is opposite to the work done by specific lift $\cal{L}$ in making displacement $h$ at the rate of $h_t$, $E$ is opposite to the work done by $\cal{L}$ at the rate of the $z$-component velocity $W$.  From these integral expressions defining $P, E, T$, it is obvious that they are related by
\be  P=E + TU  \qquad~~(\mbox{swimming energy conservation}),\label{9}\nd
automatically asserting the instantaneous conservation of energy in aquatic animal self-locomotion.
\vskip 1mm
For the ribbon plate of constant width, the added mass $m=\rho\pi b^2$ is a constant.  The integrals in (\ref{8}) can each be converted to give a sum of two terms, one being a time derivative and the other integrable along $0<x<\ell$.  For $E$, it is clear that
\be  E=m\int_0^\ell WDW dx=\frac{m}{2}\int_0^\ell(\partial_t+U\partial_x)W^2 dx, \quad~\longrightarrow\quad~ \overline{E}=\frac{m}{2}U\left\{\overline{W^2}\right\}_{x=\ell}, \label{10}\nd
where the time derivative of a quantity, presumed to be periodic, has the mean value zero (over a long time), yielding the integrated value at the trailing edge $x=\ell$ as the time mean of $E$, denoted by $\overline{E}$.  Similarly, it is straightforward to attain for power $P$ its mean as
\be  \overline{P}=m U\left\{\overline{h_t W}\right\}_{x=\ell}.  \label{11}\nd
Therefore, for $z=h(x,t)$, an arbitrary $C^1(x,t)$ function, $P, E,$ and $T$ have their time mean values as
\be \overline{P}=m U\left\{\overline{\frac{\partial h}{\partial t}\left(\frac{\partial h}{\partial t}+U \frac{\partial h}{\partial x} \right)}\right\}, ~~\overline{E}=\frac{m}{2}U\left\{\overline{\left(\frac{\partial h}{\partial t}+U \frac{\partial h}{\partial x} \right)^2}\right\},~~\overline{T}=\frac{m}{2}\left\{\overline{\left(\frac{\partial h}{\partial t}\right)^2}-U^2 \overline{\left(\frac{\partial h}{\partial x}\right)^2} \right\}, \label{12}\nd
where the last equation results from applying (\ref{9}), and these means are evaluated at $x=\ell$.  The foregoing presentation in this section is abstracted from Lighthill (1960, 1970) and Wu (1961, 1966).
\vskip 1mm
These basic relations are of significance in showing that when the mean squares of $h_t$ and $h_x$ in steady cruising  remain constant near the trailing edge, the energy shed into the wake, the power expended and the thrust acquired all remain constant in time.
\vskip 2mm
\noindent{\bf 2.1.1. Swimming motion parametric in bodily wave velocity} 
\vskip 1mm
Of all the general plate motion $z=h(x,t)$ of (\ref{1}), it is of interest to introduce the special case:
\be z=h(x-ct) \qquad~~~(0<x<\ell, ~|y|<b, ~t>0, ~c\in R>0), \label{13}\nd
to represent the distally progressing wave with wave velocity $c$, here abstracted as a one-parameter family, parametric in $c$, for a basic study.  By (\ref{13}), (\ref{4}) and (\ref{7}) give
\begin{subequations}\label{14}
\be W=(U-c)h_x, \qquad~~ {\cal L}=-(c-U)^2h_{xx},  \label{14a}\nd
and (\ref{12}) becomes (with the overline $\overline{(\cdot)}$ omitted)
\be  P= m Uc(c-U)h_x^2,\quad~~E=\frac{m}{2}U (c-U)^2h_x^2, \quad~~ T=\frac{m}{2}(c^2-U^2) h_x^2, \label{14b}\nd\end{subequations}
with $h_x$ evaluated at the trailing edge $x=\ell$.  Here, it is of great interest to notice that in this special case, both the mean thrust $T$ and mean power $P$ together become positive (or negative) for $c>U$ (or $c<U$).  In particular for this family, there is the hydrodynamic propulsive efficiency,
\be  \eta =\frac{TU}{P}=\frac{c+U}{2c}=\frac{P-E}{P}=1-\frac{c-U}{2c}, \label{15}\nd
which depends only on the velocity ratio $U/c$ as the sole parameter and is independent of time, after all the common factors cancel out.  Further, (\ref{15}) implies that high efficiency can be obtained throughout the time for $0<c-U\ll 2c$, ~e.g., $\eta\simeq 0.9$ for $U\simeq 0.8c$; and $\eta\simeq 0.95$ for $U\simeq 0.9c$.
\vskip 1mm
In general, fishes have other parameters, e.g. longitudinal variations in depth $b=b(x)$ and thickness, using various appended dorsal, ventral and caudal fins, some of which can even be actively controlled.  These new factors jointly result in differing performances involving additional biophysical effects such as in nonlinear and interactive ways that have attracted investigations in broad scope.  Here, we point out that the propulsive criterion in terms of the speed ratio $U/c$ retains the same standard in general as in this simple case of (\ref{15}), which is not basically affected by other generalizations.
\vskip 2mm
\noindent{\large{\bf 2.2. Bound and Free Vortex Sheets}}~
\vskip 1mm
Physically, vortex sheets result from integrating the vorticity vector across the boundary layer enveloping the body surface moving through a viscous fluid at sufficiently high Reynolds numbers (for the layer being thin).  In general, a velocity field in a boundary layer, $\mb{q}=(U+u,v,w), ~(U$ being the velocity of the irrotational flow enveloping the layer), has its vorticity field $\mb{\omega}=\nabla\times\mb{u}=(w_y-v_z, ~u_z-w_x, v_x-u_y)=(\omega_1,\omega_2,\omega_3)$, with the local $z$-axis normal to the body surface (as in (\ref{1}) for ribbon plate).  Then integrating the $(\omega_1,\omega_2)$ distributions across the layer will give the layer-integrated vorticity vector $\mb{\gamma}=(\gamma_1, \gamma_2)$, which is everywhere tangential to the layer.  Making the layer thickness vanish on reactive theory, the layer becomes a vortex sheet enveloping the body surface, of strength $\mb{\gamma}(x,y)$, with $\nabla\cdot\mb{\gamma}=\partial\gamma_1/\partial x+\partial\gamma_2/\partial y=0$ being the layer-integral of $\nabla\cdot\mb{\omega}=\nabla\cdot\nabla\times\mb{u}=0$.
\vskip 1mm
For the waving ribbon plate, integrating the vorticity in the boundary layers on the two sides of the plate accordingly yields for $|y|<b$,
\begin{subequations}\label{16}
\be && \gamma_1=-\int \frac{\partial v}{\partial z}~dz=v_- -v_+=-2W(x,t)\frac{y}{\sqrt{b^2-y^2}},\label{16a}\\[1mm]
  && \gamma_2=\int \frac{\partial u}{\partial z}~dz=u_+-u_-=\frac{\partial}{\partial x}(\phi_+-\phi_-) =-2W_x(x,t)\sqrt{b^2-y^2}, \label{16b}\nd\end{subequations}
satisfying $\nabla\cdot\mb{\gamma}=\gamma_{1x}+\gamma_{2y}=0$.  With $W(x,t)$ given by (\ref{14a}), the vortex-lines can be drawn for $|\mb{\gamma}|=c_1, c_2, \cdots$ over the plate, with each line forming a closed loop because of $\nabla\cdot\mb{\gamma}=0$ (see Fig. 1).

\begin{figure}[htb]
\begin{center}
\includegraphics[scale=0.90]{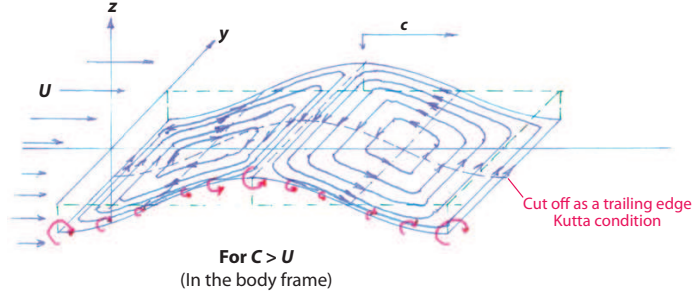}
\caption{\footnotesize The closed vortex loops on the vortex sheet bound to the ribbon plate are for plate wave speed $c$ greater than plate swim speed $U$. They are shed continuously under the Kutta condition to become free vortices drifting with the fluid in the wake. The curly arrows indicate the sense and magnitude of $\gamma_2$, the $y$-component of the vortex-sheet vector.}
\end{center}
\end{figure}
\vskip 1mm
\noindent{\large{\bf 2.3. Vortex Shedding and Wake Momentum}}
\vskip 1mm
Having found the instantaneous vortex loops on the sheet bounded to the ribbon plate, everywhere regular up to the  trailing edge, there encounters the process of shedding the vortex sheet into the wake downstream.  By physical argument, such as so clearly expounded by von K\'{a}rm\'{a}n and Burgers (1934), the flow satisfies the {\it Kutta condition} that the velocity field be regular (tangential to the body surface at the edge) and flow pressure be continuous at the trailing edge.  Once so shed, the shed vortex distribution, denoted by $\mb{\gamma}_w$, becomes a free vortex sheet and has the properties comprising (i) that {\it it satisfies the Helmholtz theorem  that it is a material property conserved with the local fluid,} i.e.
\be D_w \mb{\gamma}_w=(\partial_t+\mb{q}_w\cdot\nabla)\mb{\gamma}_w=0 \quad~~(\mbox{or}~~D_w\simeq \partial_t+U\partial_x,~~\mbox{if linearized}), \label{17}\nd
where $\mb{q}_w$ is the algebraic mean of the velocity above and below the fluid point to which the specific wake vortex $\mb{\gamma}_w$ will forever be attached.  And also (ii), it relates to the {\it circulation} $\Gamma (x,t)$ defined by
\begin{subequations}\label{18}
\be \Gamma (x,t)=\oint_C \nabla\phi (x,0,z,t)\cdot d\mb{x}=\int_0^x\gamma_2(x,t)dx=\phi_+ -\phi_-\quad~~(0<x<x_m),\label{18a}\nd
where the contour $C$ wraps around the plate from the point $\mb{x}_-=(x,0,z=0_-)$ below the sheet clockwise around the leading edge of the  plate to point $\mb{x}_+=(x,0,z=0_+)$ above the sheet, so that $\phi_+ -\phi_-$ may result with $\nabla\phi\cdot d\mb{x}=d\phi$, or $\nabla\phi\cdot d\mb{x}=(u_+-u_-)dx=\gamma_2 dx$, while $x_m~(>1)$ is the location arrived at by the starting vortex at time $t>0$.  Then we have (ii), by {\it Kelvin's theorem},
\be \Gamma (x,t)= \int_0^x \gamma_2(x,0,0,t)dx=\Gamma (x,0)=\int_0^x \gamma_2(x,0,0,0)dx=0\quad~~(x>x_m),\label{18b}\nd \end{subequations}
stating that $\Gamma (x,t)=0$ for $x>x_m$ since $\phi_+ =\phi_-$ there (with no vortex sheets), if $\Gamma$ is initially zero.  This principle is described here for the mid-span of the ribbon plate to also include the 2D case.
\vskip 1mm
For the ribbon plate waving from at rest, vortex shedding occurs at time $t=0+$, since solution (\ref{4}) holds at once over the entire plate, with a $y$-component velocity jump of $\gamma_1=v_--v_+$.  From (i)-(ii) it follows that the vorticity shedding is continuous at the trailing edge so that behind the trailing edge there is a new stretch of free vortex sheet with the same jump in ($v_--v_+$), because otherwise, any discontinuity in ($v_--v_+$) would give rise to a singular force impossible to exist.  Being free, the sheet drifts, by (\ref{17}), with the local fluid of the same velocity, hence no force acting on it, implying the pressure continuous across the sheet (in sharp contrast to the bound vortex sheet on the plate in bearing force).  More specifically, as the wake gains a circulation $\delta\Gamma$ as the body moves forward a short distance $-U\delta t$ in the absolute frame, the plate loses, by (\ref{18b}), a circulation by $-\delta\Gamma$.  Furthermore, as the lateral momentum $mW$ being shed with a gradient $mW_x$, the free vortex sheet is rotated about the $y$-axis at the rate of $\partial W/\partial x$, with the angular velocity $-\partial W/\partial x$, together with the $z$-momentum $mW$ also rolled into the core of the rolling vortex.  This causes a loss of the $x$-component of momentum by $-mWW_x=-(mW^2/2)_x=-\partial E/\partial x$, which accounts for the energy conservation in the wake.
\vskip 1mm
For the waving ribbon plate, the wake vorticity $\gamma_2$ (proportional to $-W_x$ by (\ref{16b})) reaches its maximum when the tail (trailing edge) reaches its maximum displacement, lending the free vortex sheet to roll at its maximum rate to form the center of a free rolling vortex in counter-clockwise sense.  Similarly, a rolling vortex forms when the tail reaches its minimum displacement, producing a vortex rolling in the clockwise sense.  The two rows of trailing vortices clearly pumps the fluid in between to gain an excess of momentum going downstream (along a bit undulatory path), thus affording a forward thrust needed to overcome the drag acting on the body by the viscous fluid (cf. Fig. 2).  This vortex street may be called the {\it locomotive vortex street}, in sharp contrast to the famed von K\'{a}rm\'{a}n `vortex boulevard' behind a blunt body being dragged by the viscous fluid.  This picture of the wake vortex assembled with the waving-pate then accounts for the conservation of momentum in its entity.
\begin{figure}[htb]
\begin{center}
\includegraphics[scale=0.88]{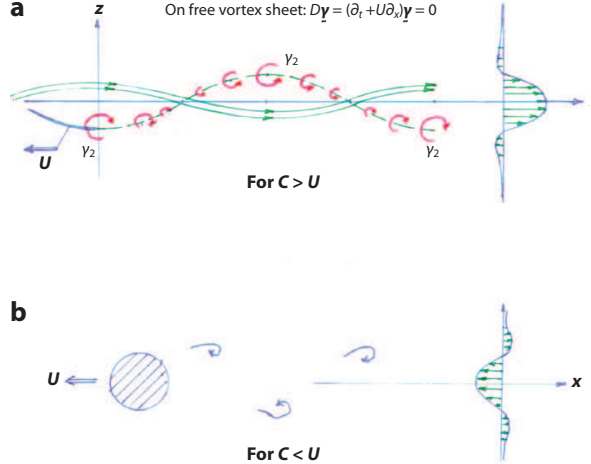}
\caption{\footnotesize The {\it locomotive vortex street} for $c>U$ in the laboratory frame, sending excess momentum downstream and giving thrust to the plate (cf. K\'{a}rm\'{a}n \& Burgers 1934, Fig.105; Lighthill 1969, Fig.8). All the arrows reverse in sense for $c<U$, and drag on the body.}
\end{center}
\end{figure}
\vskip 0.2mm
\noindent{\bf 2.4. Generalizations.}
\vskip 1mm
Aside from the simple case expounded so far, it requires various generalizations to cover most of the remaining main factors involved in fish locomotion.  Some principal factors raise such issues as the body thickness, variations in body depth, $db(x)/dx$, vortex shedding from dorsal, ventral, and caudal fins, with effects on body movement, and {\it vice versa}.
\vskip 1mm
In addition, of the various species of fishes, there are distinct modes of swimming movements encompassed by fluid mechanical studies on e.g. the carangiform and thunniform on one hand, and those in balistiform and gymnotiform on the other.  In the carangiform mode, body undulation becomes increasingly more amplified only in the posterior half, or in an even shorter distal end.  The mode of balistiform and gymnotiform describes a variety of species of fishes that propel themselves by passing a wave along an elongated dorsal and ventral fins with the body held nearly straight.  This latter type of fish, as told, have been observed to dash out swiftly from a straight tunnel, grabbing their preys, and then back straight (without turning) into the tunnel (not wide enough for body to wave), using only their elongated fins.  Typically, all fish bodies may be regarded as slender.
\begin{figure}[htb]
\begin{center}
\includegraphics[scale=0.72]{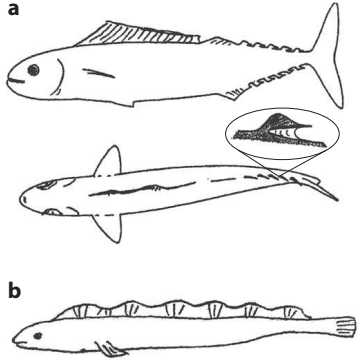}
\caption{\footnotesize (a): Two fish forms show dorsal and lateral fins of mackerel ({\it Scomber scomber}) with a row of caudal finlets with an operable membrane.  (b): A view of balistiform and gymnotiform locomotion by propagating waves along the elongated dorsal fin with the body stretched straight.}
\end{center}
\end{figure}
\vskip 0.2mm
\noindent{\bf 2.4.1. Body thickness effect.} ~In general, fish body has a symmetry about its center-plane, which may be called the {\it focal plane}, across which the body is thin.  The irrotational flow past such a body in undulation of small amplitude can be evaluated with a distribution of flow singularities (mass source-sinks and dipoles) at the focal plane for construction of accurate, even exact solutions by applying a scheme known as the {\it focal-plane singularity method}, developed for general case by Wu \& Chwang (1974), Miloh (1974) and others\footnote{This {\it focal-plane image method} ~can be exemplified by the irrotational flow of a uniform stream past a sphere represented by a certain distribution of a source-sink pair on the front and rear half of the sphere, or a similar pair of stronger magnitude on a smaller confocal sphere, or ultimately by a dipole at the focal point at the sphere center.}.  This scheme has been quite actively applied in the naval hydrodynamics profession for studies on ship maneuvering and see-keeping in ocean waves, even in designs of bulbous bow extending forward from the keel for reducing the self-induced water wave resistance.
\vskip 1mm
In fish propulsion, the body thickness is found to be secondary in effect by Wu \& Newman (1972), which can be neglected, or amended if desired, as shown by Newman (1973) and Newman \& Wu (1973) who employed conformal mapping to map the cross-sectional body contour into a plate for drawing the conclusion (see also Yates, 1977; 1983).  This also agrees with
Lighthill (1975) in giving the added mass as $m=\beta\rho\pi b^2$ parametric in $\beta$, which varies little from unity.
\vskip 1mm
However, accounting for body thickness is useful to analyzing the effects due to vortex shedding by undulatory dorsal and ventral fins such as in studies on the balistiform and gymnotiform modes by Wu (1984) and Lighthill \& Blake (1990a,b). In these cases, the bodily motion will vary with both $x$ and $y$, as to be discussed in \S 2.5 and \S 2.6.
\be  z~=~h(x,y,t) \quad~~ (x,y\in S_b|~0<x<\ell, ~|y|<b(x)). \label{19}\nd
\vskip 2mm
\noindent{\bf 2.4.2. ~Vortex shedding from appended fins.} ~This is an important feature on vortex shedding from appended dorsal, ventral and caudal fins in fish swimming.  For simplicity, we consider the simpler case of locomotion by planar undulation, i.e. $z=h(x,t)$, independent of $y$ as in (\ref{1}) and with body thickness still neglected here, but with side fins appended to body of varying depth.  This generalization can be implemented by dynamical arguments for the anterior, posterior and caudal sections.
\vskip 2mm
\noindent{\bf (A)}{\it The anterior leading-edge section} ($0 < x < x_e, ~db/dx =b'(x)>0)$
\vskip 1mm
In the leading portion anterior to the widest body section (at $x = x_e$) of fish body of increasing depth with no fin, the lateral swaying movement is analogous to the ribbon plate.  It pushes the water, giving the transverse slice of water a $z$-component momentum per unit length in $x$ as in (\ref{6}),
\be {\cal M} = m(x)W(x,t)\quad~~ W(x,t)=Dh\equiv\partial h/\partial t+U\partial h/\partial x\quad~~(|y|<b(x)).\label{20}\nd
where $m(x)=\rho\pi b^2$ is the added mass per unit length, now a variable.  In analogy with (\ref{7}), the instantaneous specific lift, ${\cal L}$, equal and opposite to $D{\cal M}$, is generalized to
\be {\cal L}(x,t) =-D{\cal M}=-(\frac{\partial}{\partial t}+U\frac{\partial}{\partial x})[m(x)W(x,t)]\quad~(0<x<x_e).\label{21}\nd
This can be used to deduce, in analogy with (\ref{8}), the power, thrust, and energy for this section.
\vskip 1mm
Furthermore, the total pressure acting on this well-rounded forward body can produce a thrust, which has been evaluated by Wu (1961)(1971a,d) along a simple approach treating the body as a focal plane with a dipole distribution for the slender-body scheme.  The thrust produced due to the cross flow across the body section at station $x ~(0<x<x_e)$  consists of two parts, $T_p+T_s$, with $T_p$ given by the regular pressure on the plate and $T_s$ the thrust due to the singular pressure at the sharp leading edges of the focal plane, as given by Wu (1971a) on plane airfoil theory.
\vskip 2mm
\noindent {\bf (B)}{\it Trailing side-edge section} $(x_e< x < x_c, ~db/dx =b'(x)<0)$
\vskip 1mm
Posterior to the body section of maximum span ($x>x_e$), various species of fish (such as dolphin fish {\it Coryphaena hippurus}, yellowtail {\it Seriola quinqueradiata}, mackerel {\it Pleurogrammus azonus}, mullet {\it Mugil cephalus}, etc.) have dorsal and ventral fin edges slanted rearward as trailing edges from which the flow leaves smoothly, under the Kutta condition, to form vortex sheets trailing along the body side, up to the caudal peduncle at $x=x_c$ from which the caudal fin follows.  Under this condition, the effect of shedding vortex sheets along body side was found by Wu (1971b,c,d) to make the material rate of change in fluid momentum independent of the longitudinal variation of the cross-flow added mass (due to body shape alone) in calculating the lift, thus yielding the specific lift in this section as
\be  {\cal L}(x,t) = -m(x)~DW(x,t) \quad~~~ (x_e <x< x_c), \label{22}\nd
where $m(x)=\rho\pi b^2(x)$.  This result shows that {\it the effect of vortex sheets shed from side-fin trailing edges alongside a body section is to have the momentum transported as if $m(x)$ is a constant}, although $m'(x)< 0$.  Physically, this implies that the vortex sheets fill in the gap left by the reduced body plate  width, making the new flow boundary like a ribbon plate uniform in span (of width $b_e=b(x_e)~\forall ~x>x_e$).  We also note that there is no discontinuity between the lift ${\cal L}(x_e-0,t)$ of (\ref{21}) and the lift ${\cal L}(x_e+0,t)$ of (\ref{22}) provided $b(x)$ and $W(x,t)$ are both $C^2(x=x_e)$ and $b'(x_e)=0$.
\vskip 1mm
Lighthill (1970) considered another morphological configuration with the main dorsal fin terminated abruptly with a straight transverse trailing edge at $x=x_s ~(x_e <x_s < x_c)$, from which a vortex sheet is shed to fill the region between the dorsal and caudal fins (such as in species {\it Cyprinnidae} and cat-fish Siluroidea).  Lighthill's theory shows that in addition to the vortex-free added mass $m(x)$ of a section $S_x$ at $x$ ($x_s <x < x_c$), there is also the added mass $\widetilde{m}(x)$ due to the frozen body segment, giving
\be {\cal L}(x, t) = - D\{m(x)W(x,t) + \widetilde{m}(x) w(x_s, \tau_f)\} \qquad (x_s <x < x_c), \label{23}\nd
where $w(x_s, \tau_f)$, being stemmed from (\ref{17}) for material conservation of free vortices, denotes the value of the $z$-component velocity $w$ depending on the fin's trailing edge at $x_s$ and on the '{\it retarded time}' $\tau _f = t-(x-x_s)/U $, which is the time when the water slice was leaving the fin's trailing edge at $x_s$ and reaching the section $S_x$ (located at $x$) at time $t$.
\vskip 2mm
\noindent {\bf (C)}{\it The caudal-fin section} $(x_c<x<\ell)$
\vskip 1mm
In this section, the interaction between the reentrant vortex sheet and the caudal fin has been argued as being capable of augmenting handsomely the lift and thrust generation, especially when they become opposite in phase.  The problem of evaluating this dynamic interaction is however complicated.  An earlier attempt by Wu (1971b) contains a deficiency which was removed by Wu \& Newman (1972) who took, in addition,  the body thickness effect into account, yielding a solution to the caudal fin problem in terms of the Abel integral equation which can be evaluated numerically.  A consistent slender-body theory has been subsequently developed by Newman \& Wu (1973), Newman (1973),  Newman \& Wu (1975) by taking into account both the kinematic and dynamic interactions and the effect of body thickness on trailing vortices.  This problem will be discussed in \S 2.5 for the generalized case of a thin body-fin system moving with displacement $h(x,y,t)$ varying with both $x$ and $y$.
\vskip 1mm
To this end, we remark that for fishes known for their high performance using the so-called 'lunate', or crescent-moon-shaped tails for propulsion, it calls for abandoning slender caudal fin analysis in favor of adopting high aspect-ratio oscillating wing theory.  Earlier advances can be referred to K\'{a}rm\'{a}n and Burgers (1934) who showed that high propulsive efficiency cannot be achieved from either the pitching or the heaving mode alone, but with the two modes superposed with an optimal phase lag.  
\vskip 1mm
In this respect, it is ingenuous of Lighthill (1970) to introduce a single parameter he called the '{\it proportional feathering}' (defined as the product of the maximum incidence angle of a flapping wing and the ratio of the flight velocity to wing's maximum tranverse velocity), or $\vartheta=U\alpha/h\omega$ in the same notation as above.  For heaving and pitching modes, $\alpha=kh$, $k$ being the wave number, hence $\vartheta=U/c$, implying, by (\ref{15}), that $1-\vartheta\ll 1$ should be ideal for feathering.  Physically, this illustrates how the thrust making and its rate of working in the total power expenditure can vary in a competing trend after putting geometry, physical principles and simple yet powerful mathematical methods together in formulating the control problem to attain valuable results and draw lucid conclusions.  It is with this clear feathering concept that the 2-D lunate tail wing theory was first expounded by Lighthill (1970).  This feathering parameter has been effectively used by Chopra (1976) to extend the lunate-tail theory to oscillations of large amplitude (for wings of infinite aspect-ratio) and by Chopra \& Kambe  (1977) to predict the hydrodynamical behavior of a variety of lunate-tail planeform of large aspect-ratio.   
\vskip 2mm
\noindent {\bf 2.4.3. Sudden start and change in direction of fish locomotion}
\vskip 1mm
We note that the line of approach just elucidated is also applicable to fish motions with a sudden acceleration or change in direction, as found by Wu (1962) for a waving flexible plate swimming with a rectilinear arbitrary speed $U(t)$.  For acceleration starting from  rest, the extraneous power $P$ is found positive definite at order $O(t)$, whereas the thrust can arise only from $O(t^2)$; after that the optimum motion form can be determined for maximum thrust with power fixed.  The problem of fish turning was studied by Weihs (1972) for a gold fish turning $90^\circ$ in a distance of half its length with practically no loss in kinetic energy, a feat much beyond any small perturbation scheme.  This highly efficient motion exhibits itself with the evidence that the locus of the fish centroid appears to be precisely the envelope of all the images of the fish centerline (or backbone).  Similarly, the problem of fish starting from rest was analyzed by Weihs (1973) with the fish centerline set in a preparatory gentle $C$-shape to send a straightening whip of the distal body end to send the fore body dashing forward without turning.  These two works were discussed extensively in depth by Lighthill (1973a).
\vskip 1mm
More recently, this fast-force production problem has been further studied analytically and numerically using a tail-flapping model by Hu et al.(2004).  Both the $C$-start for avoiding predators and the $S$-start for catching preys rely on placing the backbone in a preparatory $C$ or $S$ shape for shedding an impulsive tail vortex as found to be important to impulsive thrust generation.  The efficiency can be optimized by taking larger deflections with no delay at stroke reversal.  An integrated method has been developed by Yang et al. (2008) to study 2-D self-propelling aquatic and aerial locomotion involving unsteady body deformation dynamics, obtaining results in close agreement with experiments.

\vskip 2mm
\noindent {\bf 2.5. A generalized slender-body theory of fish locomotion}
\vskip 1mm
The general case of body-fin movements that vary with both $x$ and $y$, as that found in balistiform and gymnotiform, has been investigated by Wu (1984) and also by Lighthill \& Blake (1990a,b).  This can be taken as a boundary-value problem based on the linear slender-body theory, with the flow represented by a vortex distribution on the body focal plane or using conformal mappings.  This irrotational flow in a fluid incompressible and inviscid has the velocity potential $\phi_o (x,y,z,t)$ such that
\be  \phi_o = Ux + \phi (x,y,z,t)\quad~~ (\nabla^2\phi_0 =\nabla^2\phi=0), \label{24}\nd
giving the perturbation velocity (relative to the mean body frame of reference) as
\be \mb{u}=(u,v,w)=\nabla ~\phi \qquad ( \nabla =(\partial_x,\partial_y,\partial_z)). \label{25} \nd
\vskip 1mm
For slender bodies of length $\ell$ and depth $b$ characterized by slenderness $\delta=b/\ell\ll 1$, we can scale $y,z$ by $b$ and $x$ by $\ell$, thus converting the equation $\nabla^2\phi=0$ in (\ref{24}) under the new scales to
\be   \phi_{yy} + \phi_{zz}+\delta^2\phi_{xx}  = 0 . \label{26}\nd
If the body motion displacement is also of order $z=h(x,y,t)=O(\delta\ell)$, (\ref{26}) then provides the foundation for developing the slender-body theory starting with the expansion $\phi=\phi_0 +\delta^2\phi_1+\cdots$, which produces, by substitution in (\ref{26}), a set of ordered equations for $\phi_k ~(k=0, 1, 2, \cdots)$ as
\be \phi_{yy}+\phi_{zz}=0\quad~(k=0),\quad~~\phi_{(k)yy}+\phi_{(k)zz}=\phi_{(k-1)xx}\quad~~(k=1,2,\cdots).\label{27}\nd
In practice, the leading order may suffice unless $h$ is of large amplitude, as we will discuss in \S 2.6.
\vskip 1mm
In the leading order, the Bernoulli equation for fluid pressure $p$ can be linearized to give
\be D\phi =\Phi,\quad~~\Phi =(p_{\infty}-p)/\rho\qquad (D\equiv\partial/\partial t+U\partial/\partial x),\label{28}\nd
where $\Phi$ is {\it Prandtl's acceleration potential}, $\rho$ the fluid density.  From ({27}) and ({28}) it follows that $\Phi$ is also harmonic, with its conjugate function $\Psi$.  In terms of complex variables as used by Wu (1971b, 1984), the coordinate $\zeta$, velocity potential $f$, acceleration potential $F$, and complex velocity $\chi$,                     \be \zeta=y+iz, \quad~~  f=\phi +i\psi, \quad~~ F=\Phi +i\Psi, \quad~~ \chi=v-iw={df/d\zeta}, \label{29}\nd
are analytic functions of one another and are related by
\be  Df = F \quad~~~ D\chi ={dF/d\zeta}\quad~~ (D\equiv{\partial /\partial t}+U{\partial /\partial x}).\label{30}\nd
The boundary conditions for this problem can be expressed in terms of either $\phi$ or $\Phi$ as follows.
\vskip 1mm
\noindent I. Boundary conditions in terms of $\phi$ \qquad\qquad~~II. Boundary conditions in terms of $\Phi$\qquad
\be
(i) \quad ~~ (\phi_z)_\pm &=& Dh = W(y;x,t),  \qquad\quad ~~ (\Phi_z)_\pm = DW(y;x,t)  \qquad ((x,y)\in S_b),\notag\\
(ii)\qquad D\phi_{\pm} &=&0,\qquad\qquad\qquad\qquad\qquad ~~~~\Phi_{\pm}=0,\qquad\qquad\quad~~~~((x,y)\in S_w),\notag\\
(iii)\qquad~~~\phi_{\pm}&=&0,\qquad\qquad\qquad\qquad\qquad~~~~\Phi_{\pm}=0,\qquad\qquad\qquad~((x,y)\in S_c),\label{31}\\
(iv)\qquad D\phi_{\pm} &=&0,\qquad\qquad\qquad\qquad\qquad~~~~\Phi_{\pm}=0,\qquad\qquad\quad~~~~((x,y)~\in~T.E.),\notag\\
(v)\qquad\quad~ f~ &=& O(\zeta^{-1}),\quad F = O(\zeta^{-1}),\quad \chi=O(\zeta^{-2})\qquad~ (|\zeta|\rightarrow\infty ).\notag\nd
\vskip 0.2mm
\begin{figure}[htb]
\begin{center}
\includegraphics[scale=0.52]{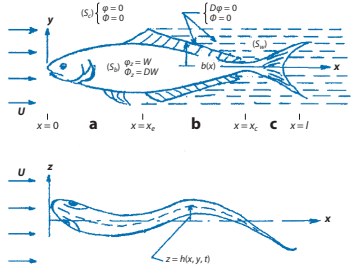}
\caption{\footnotesize The various sections of flow regions for analyzing fish propulsion: (A) anterior leading-edge section ($0<x<x_e$), (B) trailing side-edge section ($x_e<x<x_c$), and (C) caudal fin section ($x_c<x<\ell$).}
\end{center}
\end{figure}
\vskip 0.2mm
Here, the kinematic conditions I(i) and II(i) over the body planeform $S_b$ at $z=0$ prescribe the fish body lateral sway with velocity $W(y;x,t) ~(\forall~|y|<b(x))$.  In (ii), $S_w$ is the wake planeform resulting from projecting the vortex sheet shed from trailing edges of side fins onto the $z=0$ plane.  In (iii), $S_c$ denotes the region at the $z=0$ plane which is complementary to $S_b+S_w$; this condition is due to the symmetry in (i).  The Kutta condition (iv) requires the pressure to remain continuous at the trailing edge (T.E.) of side fins and caudal fin (with the pressure $p$ gauged by (\ref{2b}) as in (ii) and (iii)).  And the asymptotic behavior in (v) stems from Kelvin's circulation theorem so that the flow solution be source free and vortex free at $\zeta =\infty$.  Schematically, this is shown in Fig. 4.
\vskip 1mm
The problem formulated above is in a form of the Riemann-Hilbert problem.  By classical analysis, the required solution can be obtained for different longitudinal sections $S_x$ of the body as follows.
\vskip 2mm
\noindent (A) The anterior leading-edge section ($0 < x < x_e, ~db/dx =b'(x)>0)$
\vskip 1mm
In this section, the side edges along $y=\pm b(x)$ are assumed well rounded and moving well feathered so that the velocity field of the plane flow at station $S_x$ can be represented by a distribution of vorticity $\gamma (x,y,t)$ on line $L$ at body section $-b<y<b$ at $z=0$, i.e.
\be \chi=v-iw=\frac{1}{2\pi i}\int_L ~\frac{\gamma(x,y_1,t)}{y_1-\zeta}dy_1\qquad~~(\zeta=y+iz \notin S_b).\label{32}\nd
For condition $I(i)$, we let $\zeta\rightarrow z\pm i0 ~(|y|<b)$ on $S_b$ and apply Plemelj's formula, giving
\be v_{\pm}-iw_{\pm} = \pm\frac{1}{2}\gamma (x,y,t)+ \frac{1}{2\pi i} P\int_L ~\frac{\gamma (x,y_1,t)}{ y_1-y} dy_1 \quad~~ (y,y_1 \in S_b),\label{33} \nd
where the integral designated with $P$ assumes Cauchy's principal value.  Hence, $\gamma (x,y,t)=v_+-v_-$, and by $I(i)$, we have the integral equation for $\gamma (x,y,t)$ given as
\begin{subequations}\label{34}
\be W(x,y,t)=w_{\pm}= \frac{1}{2\pi} P\int_L ~\frac{\gamma (x,y_1,t)}{ y_1-y} dy_1\equiv G[\gamma (x,y,t)]\qquad~(|y|<b),  \label{34a}\nd
where $G$ denotes the integral operator as shown, and we have the solution ($P$ for $\int$ being omitted) as 
\be \gamma (x,y,t)=-\frac{2}{\pi}\int_L~\frac{H(y,b)}{H(y',b)}\frac{W(x,y',t)}{y'-y}~dy'\equiv G^{-1}[W(x,y,t)]\quad~~(H(y,b)=\sqrt{b^2-y^2}), \label{34b}\nd
where $H(y,b)=\sqrt{b^2-y^2}$ is the complementary solution with the symmetry (i) of (\ref{31}), and $G^{-1}$ is the inverse operator to $G \ni G^{-1}G=G G^{-1}=1$.  This is readily verified, for substituting (\ref{34b}) in (\ref{34a}) results in an identity for $W(x,y,t)$ by virtue of Poincare-Bertrand's formula:
\be  \int_L\frac{dy_1}{y_1-y}\int_L~\frac{f(y_1, y')}{y'-y_1}dy'=-\pi^2 f(y,y)+\int_L dy'\int_L~\frac{f(y_1, y')}{(y'-y_1)(y_1-y)}dy_1. \label{34c}\nd\end{subequations}
Therefore, the velocity field in this section results from (\ref{32}) and (\ref{34b}) as
\be \frac{df}{d\zeta}=\chi (\zeta)=v-iw= \frac{1}{2\pi i}\int_L~\frac{dy_1}{y_1-\zeta}G^{-1}[W(x,y,t)]\qquad (0<x<x_e). \label{35}\nd
\vskip 2mm
\noindent (B) Trailing side-edge section $(x_e< x < x_c, ~db/dx =b'(x)<0)$
\vskip 1mm
Within this section, the slant elongated dorsal and ventral fins are trailing edges (with $b'(x)<0$), and with the fins waving in displacements that may vary also with $y$, the Kutta condition being a prerequisite.  Under this condition, we note that there is a complete analogy between the two sets of boundary conditions in I(i) on  $\phi_z = Re (df/d\zeta)=W$ and II(i) on $\Phi_z = Re(dF/d\zeta)=DW$, other conditions being equal. Whence, by analogy with (\ref{31}), we attain for $dF/d\zeta$ the solution:
\begin{subequations}\label{36}
\be  \frac{dF}{d\zeta}=\frac{1}{2\pi i}\int_L~\frac{dy_1}{y_1-\zeta}G^{-1}[DW(x,y,t)] \quad~~(x_e<x<x_c),\label{36a}\nd
where operator $G^{-1}$ is given in (\ref{34b}).  Upon integration, we obtain $F$ under condition (v) as
\be F(\zeta;x,t)=-\frac{1}{\pi^2i} \int_{-\infty}^{\zeta}{d\zeta}\int_{-b}^b\frac{dy_1}{y_1-\zeta}\int_{-b}^b\frac{H(y_1)}{H(y')}\frac{DW(x,y',t)}{y'-y_1} dy'  \label{32b} \nd\end{subequations}
which is seen to satisfy the Kutta condition in particular, since $F$ is continuous everywhere, including at fin's trailing edges at $y=\pm b(x)$, by (iii) in (\ref{31}).  This $F$ field provides the pressure distribution within this trailing-side-edge section.  The corresponding $f(\zeta)$ can be obtained, straightforwardly by integrating (30), i.e. $D f=f_t+U f_x=F$, along the linearized characteristic lines $x-Ut = \xi$ = constant, starting from $x=x_e$ (at which $f$ is known from (\ref{31})), with $F$ given by (\ref{32b}).  Then the velocity field can be deduced from  $\chi =df/d\zeta$, thus yielding the $(\pm)$-side limit of $\chi =v -iw$ over $S_b$ and $S_w$ (for the details, see Wu, 1984).  The result of $w_{\pm}=w(y,\pm{0};x,t)\equiv W_v(y;x,t)$ gives the '{\it sidewash}' induced by the entire vortex system consisting of the bound vorticity lying in the planar body surface $S_b$ and the free vortex sheet over $S_w:(b(x)<|y|<b_e,~x_e<x<x_c)$.  This induced velocity field, when convected with the free vortex sheet into the caudal fin section, is responsible to assuming an important role in interacting with the caudal fin to enhance its thrust production and propulsive efficiency (cf. Wu 1984).
\vskip 1mm
The merits of using the acceleration potential $\Phi$ are clearly exhibited here in the analysis showing that $\Phi$ vanishes on all wake $S_w$ and complementary $S_c$ regions including, by (iv) of (\ref{31}), the trailing edges.  Only when the velocity field is finally attained by integration along the characteristic lines do the velocity jumps appear across the free vortex sheets on $S_w$ and with velocity being continuous on $S_c$.
\vskip 2mm
\noindent (C) The caudal-fin section $(x_c<x<\ell)$
\vskip 1mm
In this section the caudal fin has new leading edges in presence of the vortex sheets incident from upstream which in turn interact with the fin motion.  In the framework of linear theory, the velocity field can be constructed by superposition of two components, one being $\chi_v$ which is that induced by the entire vortex sheets and is convected intact downstream (with velocity $U$) into the caudal fin section, and the other, $\chi_c$, which is the new velocity due to the motion of the caudal fin itself, giving
\be \chi (\zeta;x,t) = \chi_v (\zeta;x,t) + \chi_c (\zeta;x,t) \qquad (x_c < x <\ell), \label{37} \nd
where $x=\ell$ marks the caudal-fin trailing edge.  Based on the known sidewash velocity, $W_v$ from (B) above, induced by the vortex sheets on the caudal-fin surface (which may be slender or a lunate tail of large aspect ratio), we can obtain for $\chi_c$ its solution in integral form (cf. Wu, 1984).

Here, we note that if $W_v$ vanishes and if the caudal fin is slender, this $\chi_c$ reduces to (\ref{35}) for the anterior section.  When $W_v\neq 0$, the interaction between the caudal-fin movement and the vortex sheet sidewash is seen possible to enhance greatly the thrust generation and propulsive efficiency, especially when their own induced velocity fields maintain opposite in phase because the two sidewash terms pertaining to $\chi_v$ and $\chi_c$ then totally augment each other.  Typical cases of numerical examples have been investigated by Su et al. (1983) and by Yates (1983), yielding results that lend quantitative support to the qualitative prediction.  We refer further details also to Wu (1984, 2001b).  Finally, we remark that if $W$ is independent of $y$, then (\ref{32}) reduces to the simpler solution as first expounded in \S 2.4.

\vskip 2mm
\noindent{\bf 2.6. Large-amplitude Fish Swimming}
\vskip 1mm
For the important problem that fish undulatory motion can be large in amplitude, Lighthill (1971, 1975) pursued this generalization with two principles.  (i) Kinematically, the trajectory of a fish moving with arbitrary amplitude can be described by using one Lagrangian coordinate, denoted here by $\xi$, initially stretched straight along the fish backbone jointly with the local cartesian system so that at time $t$ a point $\mb{x}=(x,y,z)=\mb{x}(\xi, t)$ on it is prescribed relative to an absolute frame, with fluid at rest far away, in which $\mb{x}(\xi,0)=\xi~(0<\xi<1)$ at $t=0$. The fish is assumed to move in the horizontal $y=0$ plane tangentially along a trajectory $(x(\xi,t), z(\xi,t),t)$, with unit tangent $\mb{\tau}=(\partial x/\partial\xi, \partial z/\partial\xi)$ and normal $\mb{n}=(-\partial z/\partial\xi, \partial x/\partial\xi )$, both lying in the $xz$-plane with the fish depth being always vertical.  The backbone is further assumed inextensible, i.e. $(\partial x/\partial\xi)^2+(\partial z/\partial\xi)^2=1$. (ii) Dynamically, it travels forward with cartesian velocity $\mb{u}=(\partial x/\partial t, \partial z/\partial t)$, or with tangential velocity $u$ forward and sidewise with normal velocity $w$,
\be  u=\mb{u}\cdot\mb{\tau}=\frac{\partial x}{\partial t}\frac{\partial x}{\partial\xi}+\frac{\partial x}{\partial t}\frac{\partial x}{\partial\xi}\equiv U(\xi,t), \quad~~w=\mb{u}\cdot\mb{n}=\frac{\partial z}{\partial t}\frac{\partial x}{\partial\xi}-\frac{\partial x}{\partial t}\frac{\partial z}{\partial\xi}\equiv W(\xi,t). \label{38}\nd
This normal velocity $W$ gives at the local $\xi$-station the lateral momentum per unit length equal to $mW ~(m\simeq \rho\pi b^2(\xi)), ~b$ being half the local fish-depth.  With this $mW$, it is then possible to give the instantaneous reactive force $\mb{F}=(P, Q)$ exerted by fluid on fish, after Lighthill (1975), as
\be  \mb{F}=(P, Q)=-\left[mW(-\frac{\partial z}{\partial t}, +\frac{\partial x}{\partial t})+\frac{1}{2}mW^2\mb{n}\right]_{\xi=\ell}-\frac{d}{dt}\int_0^\ell m(\xi) W(\xi,t)\mb{n}(\xi,t)~d\xi, \label{39}\nd
with $[\xi=\ell]$ indicating the value given at the caudal end.  Here, $-\mb{F}$ is the force exerted by the fish on the fluid in propulsion.  The various terms in (\ref{39}) generalize those attained in linear theory.
\vskip 1mm
This important theory can be further developed here to cover the general case with vortex shedding from appended fins moving at large amplitude also taken into account.  The underlying principle is to take a local instantaneous cartesian coordinate system $\mb{x}=(x,y,z)$ fixed at the fish centroid at time $t$, with its centroid moving forward tangentially to its backbone.  Then the local solution for the velocity potential $\phi(\zeta(\xi, t),t)$ and acceleration potential $\Phi(\zeta(\xi, t),t)$ can be derived for the anteroior and posterior sections by complete analogy with the solution (\ref{35}) for $\phi +i\psi$ and with (\ref{36}) for $F=\Phi +i\Psi$ so that the free vorticity distribution can be all evaluated, and the final solution follows by time marching.
\vskip 1mm
It is gratifying that Lighthill's classical works on fish locomotion have been nicely extended by Singh and Pedley (2008) for analyzing finite fish movements and by Cheng (1998) in developing the 3-D flow models to implement 'recoil' corrections in order to ensure zero net force and moment acting on fish body.  Jointly, a 3-D boundary-layer algorithm has also been developed using a vortex lattice method with applications to various cases for comparisons with analytical results available.  A significant step has been further ventured by Pedley and Hill (1999) in integrating fluid mechanics and fish internal mechanics for insight investigations of large-amplitude undulatory fish swimming.

\vskip 2mm
\noindent{\bf 3. BIRD FLIGHT}
\vskip 1mm
Theory of oscillating airfoils has important applications, including that applied to studies on bird and insect flight  and perhaps more in practice to aero-elasticity for expounding the fluttering forces and effects on airplane wing-fuselage system flying through air with turbulent patches and sharp gusts.  
\vskip 1mm
Of the various linear theories developed in the 1920s, Wagner (1925), in particular, first derived an integral equation of great significance to provide accurate determination of the starting vortex shed from an airfoil moving smoothly or impulsively forward.  For a 2-D airfoil $z=h(x,t) ~(-1<x<1)$ in oscillation, Wagner's integral equation reads
\be  \Gamma_0(t)+\int_1^{x_m}\sqrt{\frac{x+1}{x-1}}~\gamma_w(x,t)dx=0, \label{40}\nd
where $\Gamma_0(t)$ is the quasi-steady value of the circulation of the wakeless airfoil determined on steady airfoil theory with time $t$ frozen as a parameter, and $\gamma_w(x,t)$ is the free vorticity in the wake $(x\geq 1)$ with the starting vortex arriving at $x_m(t)$ for $t>0$.  Otherwise, these linear theories are thought to be either very complicated or hindered from being generalized, and not very transparent in physical conception.  These shortcomings have been overcome by von K\'{a}rm\'{a}n and Sears (1938) who created a simple and elegant restructuring of the bound and free vortex sheets varying on Kelvin's principle of conservation of circulation, or the moment of momentum, as will be further elucidated in \S 3.3.
\vskip 2mm
\noindent{\bf 3.1. Oscillating rigid airfoil.}
\vskip 1mm
During the early stage, rigid airfoils can oscillate in heaving and pitching as the only two modes\footnote{Separately, heaving alone has its propulsive efficiency $\eta_h$ falling with increasing reduced frequency $\sigma=\omega c/U, ~(\omega$ being the circular frequency, $c$ half the wing chord) from $\eta_h=1$ at $\sigma=0$ to its asymptote at $\eta_h=0.5$, whereas by pitching (about the mid-chord), its efficiency $\eta_p$ rises above zero at $\sigma=\sigma_o=1.781$ and increases to $\eta_p\simeq 0.4$ (Wu 1968).}.  In this respect, K\'{a}rm\'{a}n and Burgers (1934) noted that high propulsive efficiency can be achieved only with the two modes superposed with an optimal phase lag as the single parameter.  Various attempts followed to combine the two monochromatic modes with differing phase lag, such as that of seeking one phase lag for the highest propulsive efficiency with thrust specified.
\vskip 1mm
Alternatively, Lighthill (1970) proposed to take the pitching axis location as a new parameter, but with the heaving phase-lagged by $90^\circ$ behind pitching, i.e.
\be  z=h(x,t)= [h-i\alpha (x-b)]\exp(i\omega t) \qquad~(-1<x,b<1),  \label{41}\nd
where $h$ is the heaving amplitude and $\alpha$ the amplitude of pitching about the axis at $x=b$ as the new parameter.  With (41), the energy expelled to the wake was found to have a sharp minimum for $b=1/2$ (at the $3/4$-chord point) whereas the rate of working increases with increasing $b$, hence the optimal position is slightly behind the $x=3/4$-chord point (for details, see Lighthill, 1970, p.297).  This was in one of the manuscripts mutually communicated then between Lighthill and Wu for discussion and reference.  So (41) was used early by Wu (1971b) to study the tail movement of a porpoise in cruising with data reported by Lang and Daybell (1963).  The tail fluke angles relative to the path of tail base predicted by Wu (1971b) and the measured are found in excellent agreement with the pitching axis found at a mean of the $0.793$-chord point, well in the favorable range predicted by Lighthill.

\vskip 2mm
\noindent{\bf 3.2. Generalizations.}
\vskip 1mm
One important new issue is to consider wing flexibility for unsteady changes in its camber shape by command.  To explore the resulting effects on propulsive qualities, locomotion of a flexible wing was first studied by Wu (1961).  For an airfoil starting with acceleration from at rest, the unsteady effects have been examined by Wu (1962); the small time solution exhibits the decay of transient terms, enabling the optimum shape determined for the maximum thrust with fixed power.
\vskip 1mm
There are several salient features of significance in the world of self-locomotion of aquatic and aerial animals using lifting surfaces such as wings and body-appended fins.  First, the wings are in general large in aspect-ratio, a feature that would suit for an unsteady lifting-line approach.  Secondly, the periodic flapping of the wing generally involves changes in surface-shape function, e.g. from a stretched-straight wing in downward pronation  stroke to a form with a noticeable arched camber and spanwise bending in upward supination stroke.  Further, in swift maneuvering, the wings may bend and twist asymmetrically to change in orientation and trajectory, e.g. in the beautiful performance of a humming bird using a figure-eight wing flapping in hovering its body fixed in front of a flower, and then suddenly fleeting off in a flash.  All these features are so strongly nonlinear and time-dependent that a comprehensively valid theory would have to take all these factors fully into account.
\vskip 2mm
\noindent{\bf 3.3. A Fully Nonlinear Unsteady Theory for a Flexible Wing}
\vskip 1mm
Recently, a fully nonlinear and unsteady theory has been developed by Wu (2001b)-(2007) for evaluation of unsteady flow generated by a 2D flexible wing moving in arbitrary manner for modeling bird flight and fish swimming.  This is illustrated here for facilitating applications.
\vskip 1mm
Kinematically, the wing shape function and the resulting wake formation consists of a 2D flexible wing $S_b(t)$ of negligible thickness, moving with time $t\ge 0$ through the fluid in arbitrary manner, as can be specified parametrically using a Lagrangian coordinate system ($\xi,\eta$) jointly with a local cartesian system $\mb{x}=(x,y,z)$ for the movement of a flexible wing with arbitrary variations in wing shape and trajectory to produce an irrotational flow field in an incompressible and inviscid fluid.  This identifies a point $x=X(\xi, t), ~y=Y(\xi,t)$ on the boundary surface $S(t)=S_b(t) +S_w(t)$ comprising the body surface $S_b$ and a wake surface $S_w$, with $S(t)$ lying at time $t=0$ over a stretch of the $\xi$-axis (at $\eta =0$), $X (\xi, 0)=\xi ~(-1<\xi<1),~Y(\xi, 0)=0,$ and moving with time $t > 0$ in terms of the complex coordinate $z=x+iy = Z(\xi, t)$ as
\be  Z(\xi, t)= X (\xi, t) + i Y(\xi, t) \quad\mbox{on}\quad  S_b(t):(-1<\xi<1)+ S_w(t):(1<\xi<\xi_{m}),\label{42}\nd
with ~$\xi =-1$ at the leading and $\xi = 1$ at the trailing edge of the wing.  From the trailing edge a vortex sheet is being shed smoothly to form a prolonging wake $S_w$, and ~$\xi_{m}$ identifies the path $Z(\xi_{m}, t)$ of the starting vortex shed at $t=0$ to reach $\xi_m =\xi_{m}(t)$ at time $t>0$, with the fluid initially at rest in an {\it absolute} inertial frame of reference.  The flexible $S_b(t)$ is assumed inextensible ($|Z_{\xi}|\equiv |\partial Z/\partial \xi|=1$, or ~$X_{\xi}^2+Y_{\xi}^2=1, ~|\xi|<1$) and the point $\xi$ on $S_b(t)$ moves with a {\it prescribed} velocity $W(\xi, t) = U-iV$,
\begin{subequations}\label{43}
\be && W(\xi, t) = U-iV = \partial\overline{Z}/\partial t = X_{t}-i Y_{t} \qquad (|\xi|<1, ~t> 0; ~\overline{Z}=X-iY),
\quad~~\longrightarrow~~~\label{43a}\\[1mm]
  && W\partial Z/\partial \xi =(X_{\xi}X_t +Y_{\xi}Y_t)-i(X_{\xi}Y_t -Y_{\xi}X_t)=U_{s}-iU_{n}, \label{43b}\nd
\end{subequations}
$U_{s}(\xi,t)$ being the tangential and $U_{n}(\xi,t)$ the normal component of the wing velocity as prescribed, and similarly for the wake surface $S_w(t)$ for $(1<\xi<\xi_{m})$, but with its velocity derived with the solution.
\begin{figure}[htb]
\begin{center}
\includegraphics[scale=0.88]{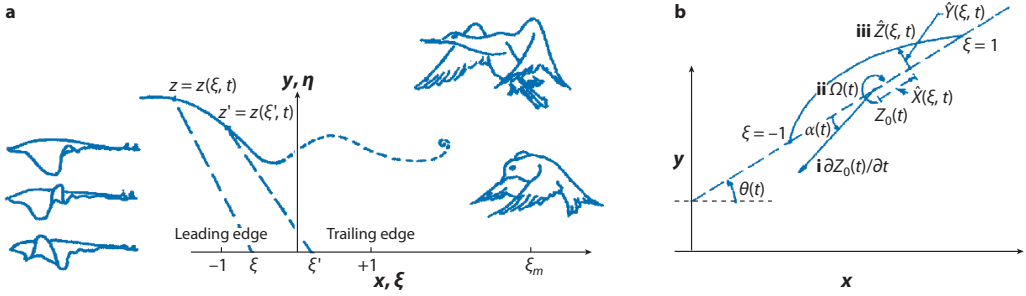}
\caption{\footnotesize (a): The Lagrangian coordinates ($\xi,\eta$) describe arbitrary motion of a 2-D flexible wing moving along an arbitrary trajectory $\eta=0$ through unbounded fluid at rest at infinity.  (b): The wing moves with (i) translational velocity ($\partial Z_o/\partial t$) at incidence angle $\alpha(t)$, (ii) rotational angular velocity $\Omega(t)$, and (iii) unsteady camber function $\hat{Z}(\xi,t)$.}
\end{center}
\end{figure}

More specifically, the wing $S_b$ moves with a {\it shape function} $Z(\xi,t)$ consisting of (i) rectilinear translation with velocity $\partial Z_0/\partial t$ at incidence angle $\alpha(t)$, (ii) rotation with angular velocity $\Omega (t)$, and (iii) flexing with {\it camber function} $\hat{Z}(\xi,t)$ which is prescribed by (see Fig. 5, the inset)
\be Z(\xi,t) = Z_0(t) + e^{i\theta}~\hat{Z}(\xi,t)\quad ~~ \hat{Z}(\xi,t) = \hat{X}(\xi,t) + i\hat{Y}(\xi,t)= \hat{X}(\xi,t) + i F(\hat{X}(\xi,t),t), \label{44}\nd
where $Z(1,t)-Z(-1,t)$ gives the chord line inclined at angle $\theta (t)$ relative to the $x$-axis fixed in the absolute frame, with pitching at angular velocity $\Omega (t)=-d\theta/dt~(>0$ for nose-up pitching).  Here, $\hat{Z}(\xi,t)$ is the complex camber function given in the body frame, with the origin set at $Z_0(t)$, corresponding to $\hat{X}(0,t)=0, ~Z_0=Z(0,t)-ie^{i\theta}\hat{Y}(0,t)$, which is the projection of point $Z(0,t)$ onto the $\hat{X}(\xi,t)$-axis, with $\hat{Y}(\xi,t)= F(\hat{X}(\xi,t),t)$ being the real camber function, and with $|Z(1,t)-Z(-1,t)|=a-(-b)=(a+b)=c(t)$ the chord length.
\vskip 1mm
For the best resolution, the simple and elegant physical concept crystallized by von K\'{a}rm\'{a}n and Sears (1938) has provided an ingenious restructuring of the vorticity distribution over $S(t)$ as follows:
\begin{tabbing}
on $S_b(t)$: \hspace*{1.2cm} \= $\gamma (\xi,t)=\gamma_0(\xi,t)+\gamma_1(\xi,t)\qquad ~~(-1<\xi<1)$, \\[1 mm]
on $S_w(t)$: \hspace*{1.2cm} \= $\gamma (\xi,t)=\gamma_w(\xi,t) \qquad\qquad\qquad ~~(1<\xi<\xi_m)$. \end{tabbing}
Here, $\gamma_0(\xi,t)$ is the vortex bound to $S_b$ in the wakeless "quasi-steady" flow past $S_b$ which can be determined by steady airfoil theory for the original prescribed $U_n(\xi,t)$, with time $t$ frozen to serve merely as a parameter for the moment.  Vorticity $\gamma_1(\xi,t)$ is the additional bound vortex induced on $S_b$ by the trailing wake vortex $\gamma_w(\xi,t)$, yet unknown, such that $\gamma_1$ and $\gamma_w$ jointly bear no change to $U_n$ over $S_b$ so as to reinstate the original time-varying normal velocity $U_n(\xi,t)$ prescribed on $S_b(t)$.
\vskip 1mm
Briefly, a cambered wing sets in motion at $t=0+$ in a uniform stream of velocity $U$ produces a quasi-steady velocity field represented by a distribution of vorticity $\gamma_0(\xi,0+)$ at $S_b(t=0+)$, i.e.
\be w(z, 0+)= u-iv={1\over 2\pi i}\int_{-1}^1\frac{\gamma_0 (\xi, 0+)}{Z(\xi,0+)-z}~d\xi \qquad~(z=x+iy \notin S_b(0+)), \label{45}\nd
where $Z(\xi,0+)$ is an initial cambered foil across which $w$ has a jump given by Plemelj's formula as
\be  u_s^{\pm}-iu_n^{\pm} = w^{\pm}(\xi,0+){dZ\over d\xi} =\pm {1 \over 2}\gamma_0 (\xi,0+)+{1 \over 2\pi i}{dZ \over d\xi}\int_{-1}^1 {\gamma_0 (\xi',0+) \over Z'-Z} ~d\xi' \quad ~(Z(\xi,0+) \in S), \label{46}\nd
with velocity $u_s$ tangential and $u_n$ normal to $S_b(0+)$ and with the integral assuming Cauchy's principal value.  Here, (\ref{46}) gives $\gamma_0 (\xi,0+)=u_{s}^+ -u_{s}^- ~(-1<\xi< 1)$ and $u_n^+=u_n^-=U_n$ normal to $S$.  With $U_n$ given on $S_b,~\gamma_0$ satisfies, for all $(\xi,\xi')\in S_b$, the integral equation:
\be U_n(\xi, t) = \frac{1}{2\pi}\int_{-1}^1 \{1+g(\xi', \xi, t)\}\frac{\gamma_{0}(\xi', t)}{\xi' -\xi}~d\xi', \quad~~
   g(\xi', \xi, t) = ~Re\left\{{dZ \over d\xi}\frac{\xi' -\xi}{Z'-Z}\right\} - 1 . \label{47}\nd
For a flat wing, the {\it residual kernel} ~$g(\xi', \xi ,t)=0$ due to $Z(\xi')-Z(\xi)=e^{i\theta (t)}(\xi'-\xi)$ for arbitrary $\theta (t)$.  For wings of small camber, solutions for $\gamma_{0}$ converges rapidly by iteration (cf. Wu 2007).
\vskip 1mm
At the same time, a vortex sheet element $\gamma_w(\xi,t_1)$ is shed into the wake of a small length $\Delta\xi=U\Delta t$ 
lying in $1\leq \xi<1+ \Delta\xi$, at this first step in time-marching calculation with $t=t_1, t_2 \cdots, t_{k-1}, t_k, \cdots (\mbox{with}~t_k-t_{k-1}=\Delta t$ ~taken aptly small).  This vortex sheet $\gamma_w(\xi,t_1)$ induces on $S_b$ a velocity distribution as $u_s^{\pm}(\xi, t_1)-i u_n(\xi, t_1)=w^\pm(\xi, t_1)dZ/d\xi ~(|\xi|<1)$, similar to (\ref{46}) for $t=0+$, of which the normal component $u_n(\xi, t_1) ~(|\xi|<1)$ has to be canceled by a new bound vortex $\gamma_1(\xi,t_1)$ on $S_b(t_1)$ in order to maintain that $U_n$ varies as specified for the plate motion.  And $\gamma_1(\xi,t_1)~(|\xi|<1)$ can be solved just like $\gamma_0(\xi,t_1)$.  Thus, the first step involves only one unknown since $\gamma_1$ is a function of $\gamma_w$.  Finally, $\gamma_w(\xi,t_1)$ can be solved by applying the Kelvin Theorem requiring that for $t=t_1, \Gamma_0 +\Gamma_1 +\Gamma_w =\int_{-1}^1(\gamma_0(\xi,t_1)+\gamma_1(\xi,t_1))d\xi+\int_1^{U\Delta t_1}\gamma_w(\xi,t_1)d\xi=0$, and similarly for $t=t_2, t_3, \cdots$ in time marching for the solution.
\vskip 1mm
Carrying out the time steps to $t=t_k ~(k=1, 2, \cdots)$ then yields an integral equation for $\gamma_w$ as
\be \Gamma_0 (t) +  \int_{1}^{\xi_m} \left\{\sqrt{\frac{\xi+1}{ \xi-1}} + N_w(\xi,t) +N_b(\xi,t)\right\}\gamma_w(\xi,t)~d\xi=0. \quad~(\xi_m=\xi(t_k), k=1, 2, \cdots). \label{48}\nd
This is Eq.(37) in Wu (2007), with the nonlinear terms $N_b(\xi,t)$ and $N_w(\xi,t)$ given therein, and is proclaimed as
the {\it general wake-vorticity theorem} in terms of which {\it nonlinear integral equation} for wake vorticity $\gamma_w$ to satisfy exactly for the solution to be exact.  It generalizes Wagner's integral equation for the linear case to account here for a flexible wing varying in arbitrary wing shape and its arbitrary trajectory.  In the linear limit, both $N_w=0$ and $N_b=0$, hence (\ref{48}) reduces to Wagner's linear integral equation (\ref{40}).
\vskip 2mm
\noindent{\bf 3.4. Comparison between theory and experiments.}
\vskip 1mm
This nonlinear theory has been applied by Stredie (2005) and Hou et al. (2007) to carry out computations of various unsteady motions of a 2D airfoil to attain results of high accuracy in all the cases pursued over a diversified broad scope.  Of them two special computational studies are presented here for comparisons with two well noted wind-tunnel experiments.  The first is for an airfoil performing heaving oscillations at reduced frequency $\sigma=\omega c/U$ ($\omega$ being the circular frequency, $c$ the half-chord, $U$ the free stream velocity) for the one-parameter family solution parametric in $\sigma$ and computed  for $\sigma=2.0$ and heaving amplitude $h=0.038$ as used in the experiment by Lai (2002).  The numerical results (for time step $\Delta t=0.001$ to $2630$ steps) is shown in

\begin{figure}[htb]
\begin{center}
\includegraphics[scale=0.86]{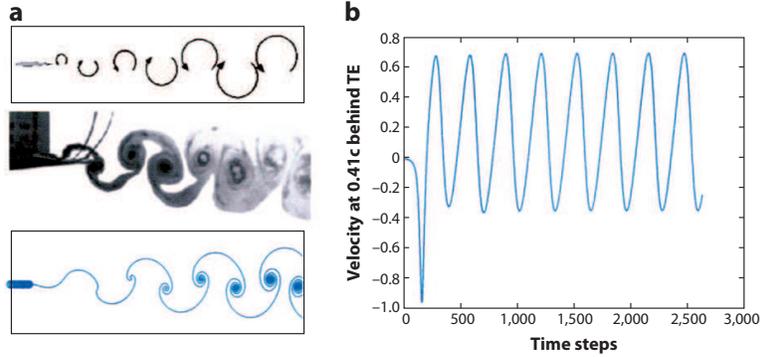}
\caption{\footnotesize (a): Heaving wing; Upper panel: rotational sense of eddies; Mid panel: experiment by Lai (2002); Lower panel: Wu's numerical results; (b): The flow velocity of a heaving wing, recorded at $0.41$ chord downstream of the trailing edge.}
\end{center}
\end{figure}

\noindent Fig. 6, with the central photo depicting the observed wake vortices shed from the trailing edge situated near the left border, growing in size with their senses indicated on the top, and with the corresponding numerical result shown in equal scale right below with a single vortex line as the centerline of the real physically diffusing vortex wake.  The qualitative and quantitative agreement between theory and experiment is excellent.  More specifically, the distance between the centers of the second pair of vortices is measured in the photo to be very nearly $0.4$ chord, so is exactly
\begin{figure}[htb]
\begin{center}
\includegraphics[scale=0.84]{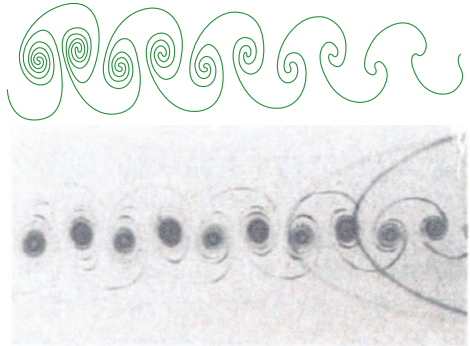}
\caption{\footnotesize A wing pitching about the $1/4$-chord point and moving from left to right; Upper panel: Wu's numerical results; Lower panel: experiment by Koochesfahani (1989).}
\end{center}
\end{figure}
the computed result.  On the right of Fig 6 is a computer plot of the longitudinal fluid velocity, in excess of $U$, produced by the heaving wing at a point $0.41$ chord downstream from the trailing edge.  The negative values (for fluid flowing upstream) in the first half period is an interesting recording of the local velocity field during the shedding of the very first starting vortex, then followed by periodic fluctuations of the local wake velocity, which is largely positive, thus implying a forward thrust exerted by the fluid on the heaving wing.
\vskip 1mm
The other case is for an airfoil performing pitching oscillations, about the quarter chord (or about the mid-cord plus heaving) computed with those data used in the experiment for the pitching amplitude of $\alpha=2^\circ$ with time step $\Delta t=0.0005$ to $3000$ steps.  Figure 7 shows the comparison of the numerical results with the experiment accomplished by Kooschefahani (1989) in which the photo of the wake vortices shed from the wing (moving from left to right) is shown below the corresponding numerical result.  The comparison is again excellent in precision.  Both comparisons stress the utmost importance in having the starting vortex computed in the very first time step in high accuracy, for a small artificial error is found to lead to growing departure of the wake vortices from the accurate results.  This also holds for all the cases examined, including mixed heave and pitch, time-periodic bending of the wing plate for simulating membrane wing (like bat flight), impulsive start in incidence and camber of a Fourier flexible wing, etc., as shown in details by Hou et al. (2007).
\vskip 1mm
The nonlinear unsteady motions of a 2D flexible wing has been extended by Hou et al. (2006), to 3D for a rectangular plate advancing normal to the flow.  These 3-D numerical results are found also in excellent agreement with the corresponding experiments (Ringuette et al. 2007).  The generalization to 3D flexible wing of arbitrary plane form could adopt some lifting-line approach for aspect-ratio no less than four, such as for the case with given spanwise free stream distribution as shown by Tuck (1975) or other types of lift-line approaches.

\vskip 2mm
\noindent{\bf 4. INSECT FLIGHT}
\vskip 1mm
The importance of studies on insect flight is evident.  The diversity in all aspects of the insect kingdom is indeed overwhelming, with close to $10^8$ species of winged insects, all self-propelling by flapping their wings of all sorts of morphological varieties, exhibiting various aerodynamical properties and distinct features as reviewed by Triantafyllou et al. (2000) and by Wang (2005).  To our mankind, one might wonder what would have become to the plant world without honey bees and butterflies, and of course there are pestering insects too.  For large insects like moths, the two wings may span to $5 cm$.  For medium to small insects down to $3 mm$, the wings are thin plane, corrugated, supported and executed by a somewhat elastic wing root to a single pair of wings (like in bee) or two pairs of forewings and hindwings (like in dragonfly), with the aspect-ratio varying from $3.8$ to $12$.  The wings of very small insects such as the various thrips can have deep fringes composed of bristles (setae) projecting outward from wing margins to form slotted sheets and to passively propagate waves generated by wing flaps so effectively that they can outperform manyfold those achievable by tiny airfoils of equal size flying at the same Re, as reported by Kuethe (1975).
\vskip 1mm
\noindent{\bf 4.1. Principal Features and Fundamental Properties}
\vskip 1mm
Kinematically, insect wing flappings can appear in various modes, e.g. in periodic pronation downstrokes and supination upstrokes, or in hovering flight, with body fixed in position, by flapping the pair of wings in repeated figure of $8$ (or figure $\infty$).  The wing pair can either keep flapping apart (without contact), or with periodic contact in the {\it clapping mode}.  For the latter case, the pioneering contribution to novel insect flight is the 'clap-fling mechanism' discovered by Weis-Fogh (1973, 1975).  This mechanism, clap to fold together and fling to fly apart, has been extensively investigated.
\vskip 1mm
Aerodynamically, there are two basic parameters in the wing flapping operation.  One is the Reynolds number, $Re=U\ell/\nu, ~U$ being a typical velocity, $\ell$ a typical length, $\nu$ the kinematic viscosity of the air, which is found to fall in a range of {\it meso Reynolds numbers}\footnote{The term of 'small (or low) Reynolds numbers' $Re$ had been commonly used, prior to the insect-flight research days, in both mechanical and biological literatures on microorganism locomotion with $10^{-6}<Re<10^{-3}$.} like $10^2-10^{4}$.  The other is the Strouhal number, $St=\omega\ell/U$, which represents the unsteady effects due to flapping of circular frequency $\omega$.  In addition, an outstanding feature in wing stroke has been observed that the insect wings in general are found to sweep at large incidence angle (of attack), like in the $20^\circ-40^\circ$ range, so much in sharp contrast to bird flight that its mechanism and effects inspire exposition.
\vskip 1mm
To this end, we may cite a famed study on the locust {\it Schistocerca gregaria} in a role-model collaboration by Weis-Fogh and Jensen (1956), a zoologist and an aerodynamicist of renown.  Placing in their wind-tunnel a specimen of weight $2g$, wing span $4 cm$, chord $c=0.75 cm$, in a wind of speed $U=4 m/s$ at $Re=Uc/\nu\simeq 2000$, they found the wing beat frequency at $f=20 Hz ~(\omega=2\pi f)$ or the Strouhal number $St=\omega c/U\simeq 1/4$, which is considered very high for the unsteady effects, as can be attested by the Stokes boundary-layer thickness $\delta=\sqrt{2\nu/\omega}$ being small (like that for a flat plate oscillating in its own plane in air at circular frequency $\omega$), which in this case gives $\delta = 0.48 mm$, a negligibly thin boundary layer on the wing.  This finding then establishes the first basic principle (i) that for $St\gg 1/Re$, the air can be assumed incompressible and inviscid for aerodynamic evaluation of lift, power and efficiency of insect flight.  This assertion should even be enhanced for smaller insects, because in general, the smaller the insect, the lower the Reynolds number, but the higher the beating frequency, hence the higher the Strouhal number.  As a result, the lift coefficient $C_L$ gained by small insects in hovering flight can be as large as from $4$ to $6$ (Weis-Fogh 1973) due to the unsteady high lift by insect wing flapping.
\vskip 1mm
An additional issue is concerned with the high incidence angles held by sweeping insect wings.  This actually is related to the clap-fling mechanism, first discussed by Lighthill (1973b), later modified by Maxworthy (1981) who discovered flow separation from the leading edge of a model wing in forming vortex sheets moving towards the wing tip as found from his laboratory experiments.  In this regard, a comprehensive view has come to focus as follows.  Ellington et al. (1996) observed flow separation around the leading edge of wings of tethered hawkmoth and a flapper model, forming a strong and stable leading-edge vortex (LEV) during downstroke to lend high lift forces.  Dickinson et al. (1999) measured unsteady forces acting on a fruit-fly wing model hovering at $Re=136, f=145 Hz$, finding out three underlying mechanisms: delayed stall in down- and up-strokes, rotational circulation and wake capture coming with rapid rotations of the wing in stroke reversal, which can provide $1/3$ of the mean lift.  More recently, Sun \& Tang (2002) numerically analyzed the 3D N-S equations for the hovering model, finding that flow separates from both the leading and trailing edges under the Kutta condition to form the leading-edge vortex plus its image (LEV\&IM) in the wing-body system and the trailing-edge vortex plus image (TEV\&IM) as the three major mechanisms.  Thus, it is clear that the high incidences held by insect wings are responsible for both (LEV\&IM) and (TEV\&IM), with the latter keeping the former stable in float, as noted by Yu, Tong \& Ma (2003).  These advances in conception have stimulated interest in insect flight to the highest level of activities in the last decade.

\vskip 1.5mm
More recently, developments have been made in aerodynamic theory and modeling, computational and experimental studies on various aspects of insect flight for our expository review below.
\vskip 2mm
\noindent {\bf 4.2. Aerodynamic Theory and Modeling of insect flight}
\vskip 1mm
It has been noted that effective aerodynamics at $Re<10^{4}$ is of great technological interest and a  fundamental scientific challenge.  Also noted, by Lian et al. (2003), the aerodynamic performance of a wing deteriorates considerably as the Re number decreases from $10^{6}$ to $10^{4}$.  For the unsteady leading-edge vortex (LEV) formation and shedding of wake vortices in insect wing flapping, Ansari et al. (2006) developed two novel, coupled, nonlinear wake-integral equations for evaluation of force and moment for 2D inviscid flows.  Separately, Bandyopadhyay (2009) regarded the effects due to the LEV phenomenon being responsible for unsteady high lift.
\vskip 1mm
A fundamental theoretical modeling for wing flapping of small insects has been developed by Yu, Tong, \& Ma (2003) for a 2D flat plate oscillating in heave and pitch, shedding vortex sheets from both the leading and trailing edges, as well as with swift rotation of the viscoelastic wing roots at stroke reversal.  Employing flow singularities of dipoles and vortices on the plate to construct the potential flow in an incompressible and inviscid fluid, results obtained have simulated the experimental conditions of Dickson \& G$\ddot{o}t$z (1993) for the unsteady lift coefficient $C_L$ at incidence angle of $\alpha =31.5^\circ$, Strouhal number $St=0.35$, which are found in good agreement.  Their $C_L$ is given in three components, one for the added mass, the other two for LEV\&IM and TEV\&IM each, exhibiting, interestingly, that the TEV\&IM is dominating in general.  Also, their results for a pitching wing are found in good agreement with the CFD results by Hamdani \& Sun (2000).  This work was revisited by Yu et al.(2005) for further expounding on the inertia effect, the LEV and TEV.  These works have been adopted by others for modeling insect flight in applications.
\vskip 1mm
In a valuable review, Tong \& Lu (2004) discussed several noticeable issues concerning insect flight.  One issue points out some shortcomings in quasi-steady schemes for unsteady insect flight that appears in need to open a new field for unsteady aerodynamics at relatively high Strouhal numbers.  Another issue stresses the fact that the pronation down- and supination upstrokes are along inclined down- and up-paths rather than the commonly assumed horizontal course, together with their marked time-duration ratios that require more attention, as revealed from the numerical studies by Liu, Ellington et al (1998), and by Sun \& Tang (2002) on 3-D N-S equations.  This last issue on stroke asymmetry has been further pursued by Yu \& Tong (2005), attaining firm evidences that the aerodynamic asymmetry between the down- and up-strokes is a key to understanding the fluid physics of generation of the lift and thrust in insect flight.  In addition, Yu, Tong \& Ma (2005) showed how the unsteady high lift can be enhanced by raising the added inertial effects, suppressing the leading-edge flow separation, and accelerating the trailing-edge  vortex separation.  Further, Bao et al.(2006) have studied the effect due to passive wing flexibility, discovering it being another mechanism for improving insect flight performance.  On this issue, Yang et al.(2008) have pursued the idea of integrating unsteady fluid mechanics and deforming body dynamics as a cross-discipline between fluids and viscoelastic solids.
\vskip 2mm
\noindent {\bf 4.3. Computational Fluid Dynamics (CFD) of insect flight}
\vskip 1mm
The problem of insect flight can be resolved numerically by CFD.  An integrated model with multi-blocked panel grid geometry has been constructed by Liu (2009) to evaluate inertial and aerodynamic forces, torques and powers expended for flapping flight in the meso Reynolds number range, with case studies demonstrating its accuracy.  The effects due to corrugation of insect wings have been found by Luo \& Sun (2005) to be small on the aerodynamic lift of fruit fly, honeybee, dragonfly, etc. ten kinds of insects, and that the effects of aspect ratio (here $2.8<AR<5.8$) are also surprisingly small.  A simple model of LEV is given by Maxworthy (2007) to explain that a spanwise pressure gradient and variations in the circulation are responsible to forming a downward propagating vortex ring.  The 3D flow past a low AR wing translating impulsively for $300<Re<500$ has been studied numerically by Taira \& Colonius (2009) for the effects of aspect-ratio, incidence angle, and planeform shape on the wake vortices and force generation, with the results validated by oil-tank experiments.
\vskip 1mm
Related to hovering of insects, Pullin \& Wang (2004) investigated the aerodynamic forces on a flat plate accelerating from rest at fixed incidence in a 2D power-law flow using a new analytically and numerically related method so specially developed.  Their results brought forth a mechanism underlying dynamic stall based on a combination of unsteady vortex lift and added mass, under a specific set of premises.  The force per unit span predicted by the vortex theory is evaluated for parameters typical of insect wings and is found to be in agreement with numerical simulations. Estimates for the shed vortices and the size of the starting vortices are also obtained, with the significance of the results deliberated as a mechanism for insect hovering flight.  More recently, the hovering of insects has been simulated by Bos et al. (2008) who applied a 2D Navier-Stokes solver to fruit fly hovering at $Re=110$ for comparisons with other models on the lift and drag, with results revealing that the forces based on their wing kinematics differ notably from those based on other simplified wing kinematics models.
\vskip 2mm
\noindent {\bf 4.4. Experimental Investigations on insect flight}
\vskip 1mm
Experimental flow visualizations and aerodynamic force calculations of three types of clap-fling mechanism have been made by Sohn \& Chang (2007).  Their tests cover the 'fling-clap-pause' and 'clap-fling-pause' two types, and their corresponding computations applied a 2D Navier-Stokes equations solver, finding good agreement between the two results for the main features about the two families of vortex pairs generated and moving off.  Results show that the leading-edge vortex bubbles are dominant features in causing a large negative pressure region near the leading edge, hence the high lift effects.  For two-pair wings of large aspect-ratio such as the dragonfly, their flight was examined by Levy and Seifert (2009) for $Re<8\times 10^3$, finding that the LEV produces relatively weak wake vorticity, thereby reducing the drag and increasing the flight performance, as is well attested by wind and water tunnel tests.  On insect hovering, Lu, Shen and Su (2007) made an electromechanical model for wind-tunnel tests, finding that (1) spanwise flow is conspicuous, (2) the main leading-edge vortex is led by a secondary dual one of the same sense, (3) development of the LEV is somewhat hampered by wing translation, and (4) the interaction between fore and hindwing pairs is detrimental to LEV development.
\vskip 1mm
\noindent{\bf 5. FISH SWIMMING AND BIRD FLIGHT IN NATURE - OPTIMUM CONTROL}
\vskip 1mm
So far, the aquatic and aerial animal locomotion has been addressed with the fluid media primarily at rest.  In wild nature with the prevailing fluids richly endowed with intrinsic energy, such as in tumbling water waves, torrential rivers, or in whirling winds, it should be of interest to explore how these animals take advantage of such energy resources by applying their talents evolved over agelong times. Intuitive observations often suggest that they are deft players in this game.  The famed view of an albatross riding on sheared wind ($U(z),0,0$), with $dU/dz>0$, gaining for its body a velocity ($u',v',w'$) in the frame of reference moving with the wind, but without doing any work.  This is accomplished mainly by taking the advantage of the work done by the inertia force at the rate of $-mw'dU/dz, ~m$ being its mass, as explained in essence by Lighthill (1974).   Also, the albatross and pelicans are seen to skim ocean waves over long distance with only gentle twisting of their wings, as expounded for the underlying mechanisms in detail by D.M. Wu et al.(2005).  Porpoises and dolphins have also been watched to ride on bow waves in front of a cruising ship with only their tail fluke aptly bent.
\vskip 2mm
\noindent {\bf 5.1. Migratory Pacific salmon}
\vskip 1mm
Another intriguing story of the Pacific Salmon returning from life-long living in the Ocean to swim thousand miles up the Columbia River to the shallow brooklets of their birth place to spawn was investigated by Osborne (1960). He tagged some migrating salmon at the Columbia Delta, and later recaptured them from fish ladders at hundred miles intervals up the river to acquire data on the time of travel and loss of body fat.  These salmon are known to stop feeding once having entered fresh water, relying on the stored fat as the sole energy supply.  He found, quite surprisingly, that a swollen river due to a torrential rain upstream did not slow those salmon down by that much of a margin (for known metabolic energy they just expended) according to frictional resistance of water being proportional to the square of the stream velocity relative to fish.  Several possible explanations were suggested, including the prospects that the salmon would be able to extract the energy of turbulent eddies for conversion into thrust, navigating through slower stream route, and other ways and means.
\vskip 2mm
\noindent {\bf 5.2. Extraction of flow energy}
\vskip 1mm
Such fascinating discoveries have inspired further studies on extraction of flow energy in general.  Earlier, the interest was first related to resolving problems of flutter of airplane wings flying in gusty air.  This requires critical consideration on the conditions of extracting energy from the stream, either actively or passively.  In the case of flutter in uniform stream, it is usually assumed that the engine maintains the flight speed constant regardless of the flutter-created inertial drag.  In waving streams, however, the flutter may become self-excited as studied by K$\ddot{u}$ssner (1935) and Garrick (1936).
\vskip 2mm
\noindent {\bf 5.3. Emulating applications}
\vskip 1mm
Generalizing wings flying in air to hydrofoils cruising in water, this important extension was led by Georg Weinblum (1954), a mentoring master in naval hydrodynamics, who developed an approximate theory for a hydrofoil heaving and pitching in regular water waves of small amplitude.  His views and conceptions, broad and profound as usual, offered foresights set on the instability and control of the naval vehicle, to both of which a clear understanding of the energy balance is of great importance.   Further studies on various related issues include that on optimum swimming shape functions by Wu (1971a,b,d), on extracting intrinsic flow energy by Wu (1972), on animals in wavy streams by Wu \& Chwang (1975), on utilizing unsteady winds and ocean currents by Wu (1980) and others.  In brief, the central idea is to find the optimum wing motion with respect to the intrinsic flow field by maintaining the feathering parameter nearly at its optimum on control theory.  For instance, there have been literatures on a family of migratory geese flying in a wedge form, led by a strong member, followed by others on the outer upwash sides.  Similarly, a school of fish also swim in diamond patterns by taking advantage of the vortices with forewash trails as so suggested in Fig. 2 and in Lighthill (1973a, Fig.12).

\vskip 2mm
\noindent {\bf 6. ENERGETICS OF FISH LOCOMOTION}
\vskip 1mm
Animal locomotion is a multi-discipline in which the frontiers of mechanics, biology and related fields converge.  It benefits from scientists working in intimate collaboration, with keen interests in continuing communication and learning from one another.  Such collaborative studies can be surprisingly rewarding, as exemplified here especially on energetics of animal locomotion.
\vskip 1mm
After the first International Symposium on 'Swimming and Flying in Nature' held at Caltech in 1974, another international symposium was organized by Lighthill and Weis-Fogh, held in September 1975 at Cambridge University on 'Scale Effects in Animal Locomotion,' a new frontier of great importance.  There, by invitation, Wu (1976) took up a new study on the scaling of aquatic animal locomotion, while emulating the comprehensive paper by Lighthill (1976) on the scaling of aerial locomotion.  It is of interest to give a recount of the proceedings then and its sequel.
\vskip 1mm
For steady fish locomotion, a mean thrust T is required of the fish to just balance the viscous drag D of water on fish, by Newton's law, in sustaining a swimming velocity $V$,
\be  T = D  \qquad~~(\mbox{force balance}).  \label{49}\nd
Here, fluid mechanics is suitable in attaining thrust $T$, by analysis and computation on reactive theory to desired accuracy as expounded in \S 2.  For viscous drag $D$, physiological schemes are suitable for evaluating the metabolic conversion from oxygen consumption by fish to muscular power with accurate results.  So, close collaborations on this joint task between mechanics and biology should be evidently fruitful, as is actually a case of splendid success.  Otherwise, it is well known that various attempts in measuring viscous drag $D$ of immobile fish or their artifical models in wind or water tunnels have produced data so much scattered as being useless (for reviews see Gray 1968; Webb 1975). 
\vskip 1mm
To proceed on the joint approach, it begins with the principle of energy conservation asserting
\be DV=P-E=\eta_m P_m ~\qquad~~(\mbox{conservation of energy}), \label{50}\nd
signifying that the animal's mean rate of working against drag $D$ is provided originally by the power $P$ expended by the fish in performing the work at swimming speed $V$ while casting off, irreversibly, some wasted energy $E$ to the fluid, with this work related to the measured net biochemical metabolic power $P_m$ consumed all for swimming (after subtracting off the basal metabolic rate which is unrelated to swimming movement), in terms of $\eta_m$ for the 'muscle efficiency' by which the biochemical energy is converted to muscular power for active swimming.
\vskip 2mm
\noindent{\bf 6.1. Metabolic rate and scale effects}
\vskip 1mm
Here, it is important to point out that the metabolic rate in fish locomotion depends on several key factors including (a) level of activity, (b) temperature of the water (affecting the concentration of dissolved oxygen), and (c) such factors as pre-conditioning (a period of fasting and exercise prior to the test to separate energy spent in digesting and interior physiological processes), as well as the state of maturity and gender of the specimen.  The level of activity is generally hard to characterize.  For the Pacific Salmon, Brett (1963, 1965a,b) has found three distinct levels of performance: the sustained (speeds that can be maintained almost indefinitely), the prolonged (speeds maintained for 1 to 2 hours with a steady effort, but leading to fatigue), and the burst (speeds achieved at maximum effort lasting for only 30 seconds).  The physiological basis of each of these levels is different and further depends on the scale of body size, such that the measured metabolic power $P_m$ assumes the general formula:
\be P_m = a~m^b, \qquad~~b = 0.775, ~0.846, ~0.890, ~0.926, ~0.970, \label{51}\nd
where $m$ is body mass, proportional to the cube of body length $\ell$, and $a, b$ are constant indices, with $b$ reported by Brett for the salmon {\it Oncorhynchus nerka}, all with an error bar of $\pm 0.145$ or less for the standard, $1/4$-max, $1/2$-max, $3/4$-max and $V_{max}$-speed, respectively.  Here $V_{max}$ denotes the speed sustainable for 60 min in fresh water at $15^o C$ and the standard state is obtained by extrapolating the $O_2$- consumption versus velocity curve to zero swimming speed.  Here we note that these $b$ values are all greater than the "surface law" ($b=2/3$) and increases with activity level to approach direct mass proportionality ($b=1$).  Brett's endeavor has made contributions of lasting value to render studies on scaling of fish locomotion more systematic and comprehensive, as we now proceed to show.
\vskip 2mm
\noindent{\bf 6.2. Scaling of swimming velocity and energy cost}
\vskip 1mm
Expressing the drag $D$ in terms of the drag coefficient $C_D$,
\be D={1\over 2}\rho V^2SC_D  , \qquad C_D =C_D(Re), \qquad  Re=V\ell / \nu  \label{52}\nd
for fish with surface area $S$ (proportional to its length $\ell$ squared) in water of density $\rho$ and kinematic viscosity $(\nu =0.0114~\mbox{cm}^2s^{-1}$ at $15^{\circ} C)$, with $C_D$ dependent on the Reynolds number $Re$ and body configuration.  From the energy balance (\ref{50}), metabolic scaling (\ref{51}), and (\ref{52}), it follows that
\be V^3 \sim \ell^{3b-2}/C_D. \label{53}\nd
For the carangiform and lunate-tail modes of swimming, both modes  being well-streamlined, Wu (1976) assumes that $C_D$ may be approximated by that for steady flow over a flat plate at equal Reynolds number, so that $C_D$ is proportional to $Re^{-1/2}$ or $Re^{-1/5}$ according as the boundary layer is laminar or turbulent, respectively, hence yielding the scaling law:
\be V=\mbox{const.} \ell^{\beta}, \quad~~\mbox{with}~~~\beta = {3\over 5}(2b-1), \quad {3\over 14}(5b-3), \quad b-{2\over 3},  \label{54}\nd
\begin{figure}[htb]
\begin{center}
\includegraphics[scale=0.75]{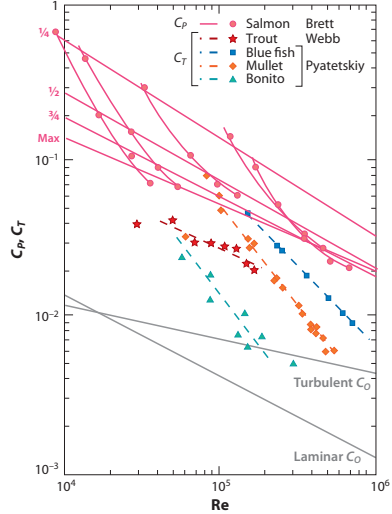}
\caption{\footnotesize Scale effects in the measured metabolic power coefficient $C_P$ for swimming exhibit variations with Reynolds numbers $Re$ for five different size groups of sockeye salmon: (solid lines): least-square-error fit to $C_P$ at specific activity levels; (dashed lines): $C_P$ variations with various activity levels for each size group; (dash-dot lines): theoretical thrust coefficient $C_T$ -- Figure adapted with Brett (1965a,b); courtesy of Academic Press.}
\end{center}
\end{figure}
\noindent according as the boundary layer is laminar, turbulent, or with $C_D=$const., respectively.  The last case (of quite large but constant $C_D$) is for the hypothetical case when flow separation occurs in the cross flow past an undulating fish body in a braking-type maneuver.
\vskip 1mm
In addition to providing the detailed metabolic rate of a single size group, a comparative physiomechanical study has been made possible by Brett's data (1965) on $P$ and swimming speed $V$ for sockeye salmon of five different size groups.  The scale effects are exhibited in the $C_P\sim Re$ relationship as shown in Figure 8 with five size groups (in dashed lines), each covering the four activity levels as marked.  The least-square-error fit for each activity level of the five size groups brings out the scale effects.  In this log-log plot of the $C_P\sim Re$ lines, the slope reads -0.61, -0.54, -0.49, -0.41 for $1/4$-max, $1/2$-max, $3/4$-max, and full level of activity, respectively, for fish over five length scales.  The result of this analysis is already an important discovery because the $1/2$-max and $3/4$-max level lines are situated across that of the reference $C_D \sim {Re}$ line for the laminar (with slope $-0.50$) in sharp contrast with the turbulent $C_D$ (with slope $-0.20$) of a smooth flat plate shown in lower Figure 8.  Also shown are the thrust coefficient $C_T$ lines derived with using the observed data of Pyatetskiy (1970) and Webb (1975).  The slopes of these $C_T$ lines are seen to be qualitatively similar to those of the estimated $C_P$ curves for fish of each scale group (see Fig. 8 ).  This scrutinizing scaling therefore brings forth the conspicuous result that {\it the viscous drag on all these specimens is primarily associated with laminar boundary-layer flows as so attested by this solid and sound evidence}.
\vskip 0.1mm
\begin{figure}[htb]
\begin{center}
\includegraphics[scale=0.68]{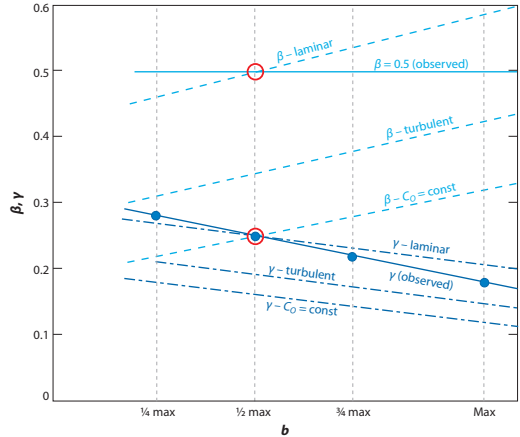}
\caption{\footnotesize Comparison of the observed values of the scale similitude $\beta$-line and the specific energy cost $\gamma$-line (adapted with Brett 1965a,b) with the predictions based on three reference states for laminar or turbulent boundary layer, and $C_D=$constant for the braking case of separated cross flows.  The $b$ values have the basal metabolic rates subtracted off, and the bio-chemically observed $\beta$-line and $\gamma$-line both intersect only with their corresponding laminar boundary-layer reference lines, and both precisely at the $1/2$-max activity level, with the intersection points circled in red.}
\end{center}
\end{figure}
\vskip 0.1mm
\vskip 1mm
In addition, we introduce the '{\it specific energy cost}', ${\cal E}$, defined for fish of weight $mg$ as
\be {\cal E} = {P_{m}\over mgV}, \label{55}\nd
a dimensionless parameter which provides a valuable measure of the relative merit of the propulsive system, signifying the energy expenditure in transporting unit weight over unit distance.  It was introduced by K\'{a}rm\'{a}n and Gabrielli (1950) in evaluating the comparative merits of fourteen classes of transportation vehicles and animal locomotion, and used by Schmidt-Nielsen (1972), Tucker (1975) and others in studies of comparative physiology, and also by Weihs (1973) in his study.
\vskip 1mm
These results for the scaling of metabolic rate and swimming velocity may now be combined to evaluate the scaling of the specific energy cost.  From (\ref{51}), (\ref{54}) and (\ref{55}) we find that
\be {\cal E}_m =P_m/mgV = \mbox{const.} m^{-\gamma}, \qquad \gamma = 1-b+\beta/3; \quad~\gamma = 0.32,~0.29,~0.25,~0.21,  \label{56}\nd
where the values for $\gamma$ are deduced by using the experimental data in Fig. 8 and the values of $b$ in (51) for the  activity level at standard, $1/4$-max, $1/2$-max, and $3/4$-max level of activity, respectively, showing how the specific energy cost $\cal E$ varies with body mass accordingly.  These values of $\gamma$ and the observed mean of $\beta =0.5$ are plotted versus $b$ (as a measure of the level of activity) in Figure 9.  In these experimental data, the basal metabolic rates have been subtracted off (cf. Wu \& Yates 1978).  Also shown are the similarity predictions of $\beta$ and $\gamma$ evaluated for the three distinct reference states of laminar, turbulent and separated flows.  Here, a comparison between the experimental data and the similitude calculations brings forth far-reaching results.  First, {\it the bio-chemically observed $\beta$-line and $\gamma$-line both intersect only with their corresponding laminar reference lines of the three flow reference states, and both at exactly the $1/2$-max activity level, with the intersection points circled in red.  This therefore shows that the boundary layers of these swimming fishes should have to be evidently laminar at these high Reynolds numbers ($8\times 10^3<Re<10^7$), and further shows that the specimens tested would have to be swimming about $1/2$-max of their maximum sustainable level}.

\vskip 2mm
\noindent {\bf 7. CONCLUSION}
\vskip 1mm
The present expository review has delineated classical studies in five major areas of animal locomotion: fish swimming, bird flight, insect flight, propulsive motion control, and locomotion energetics.  Each of these areas has marked with its own nature, outstanding properties, and significant advances.  For fish swimming, passing an undulatory wave of speed $c$ distally down its flexible slender body to swim at velocity $U$ slightly less than $c$ is the primary key feature.  To the top performers, the appended dorsal, ventral and caudal fins evolved in intricate ways display challenging nonlinear and interactive mechanisms that can now be expounded by theory and experiments developed.  Similar statements can be properly applied to bird flight.  The attractive flapping of bird wings stretched straight in downward pronation stroke and the bending and twisted wings in uplifting supination stroke can now be well modeled mathematically for analysis.  Fully nonlinear and unsteady theory can be applied for evaluation of the aerodynamics of a flexible wing moving in arbitrary manner along arbitrary trajectory for further applications.  The subject on insect flight has appeared impressive in the diversity of countless species, each with differing mode of flapping wings of unique shape and beating frequency.  The relevant aerodynamic category falls in the range of {\it meso Reynolds numbers} of $10^2<Re<10^4$.  However, the viscous effects are rendered negligible by the high values of the Strouhal number resulting from the high frequency of the insect wing flapping. In addition, recent studies in this field have discovered an outstanding characteristic mechanism rather universal in the insect world.  And this marked the wing flapping at finite incidence angles of $20^\circ-40^\circ$, causing flow separation from both the leading and trailing edges in forming a conspicuous leading-edge vortex bubble to generate unsteady high lift.  This mechanism has been adopted to make mathematical models for insect flight studies.
\vskip 1mm
On the energetics of animal locomotion, the original idea is brining the efficient and effective part of mechanics (in evaluating thrust delivered by fish) and of biology (capable of metabolic measure of muscular power for counter-balancing drag on swimming fish) for unified studies.  The results are rewarded by the invaluable evidences achieved, which would be utterly impossible otherwise.  In concluding \S 6, it has been shown, with Fig.9, that a comparison between the experimental data and the similitude predictions brings forth far-reaching results.  First, {\it the bio-chemically observed $\beta$-line and $\gamma$-line both intersect only with their respectively corresponding laminar reference lines of the three flow reference states, and both intersection points fall precisely at the $1/2$-max activity level.  These results therefore furnish the valid and sound evidences showing resoundingly that the viscous drag on fish swimming in nature is associated primarily with laminar boundary-layer flows.  And they are thus providing a valid and rigorous resolution to the `fish swim paradox' proclaimed earlier by Sir James Gray}.  That paradox has been a mechanics-biology joint stimulus that has attracted much of the diligent and dedicated studies over the decades, now bearing a fruit to both mechanics and biology disciplines for celebration.

\vskip 2mm
\noindent {\bf 8. ACKNOWLEDGMENTS}
\vskip 1mm
In taking this review of the rich literature of the classical works and new advances in this subject field, it recalls to memory all the excitement on discoveries with generations of co-workers and distinguished visiting scholars in our research group, as well as with friends and colleagues near by and far away over decades.  My warmest thanks are due to all those who have made these opportunities possible.  These interesting and challenging studies could not have been carried out by this group without the encouraging and generous sponsorships including that from The Office of Naval Research, then dynamically directed and led in this field by Phillip Eisenberg, Marshall Tulin, and Ralph Cooper, highly noted scientists themselves.  In addition, we value the same from The National Science Foundation, then ably led for this branch by George Lea, well known too for his own works.  Also deeply appreciated is the special invitation to participate in the 2000 Technion Symposium in Memory of Sir James Lighthill organized by Nadav Liron.  Further, I am greatly indebted to Milt van Dyke, Thomas Yizhao Hou, and Bing-gang Tong for inspiring and fruitful discussions.  I am also very thankful to Chien-Chung Chang, Yong Hao, Melba Bush and Chinhua Wu for valuable academic communications, literature survey, and their excellent assistance in preparation of this article.  The current work is supported in part by Fonda Wu with the American Chinese Engineering Science Foundation.
\vskip 2mm
\noindent {\bf A tribute:} ~Warmest tribute is due from me to dedicate this review article to
\vskip 0.2mm
\qquad\qquad~~ Professor Yuan-Cheng Fung, the Father of Bioengineering 
\vskip 0.2mm
\qquad\qquad in joining the international celebration of his 90th birthday.
\vskip 3.8mm
\noindent{\bf LITERATURE CITED}
\vskip 3.8mm
\noindent Ansari SA, Zbikowski R, Knowles K. 2006. Nonlinear unsteady model for insect-like flapping wings in\\ 
\indent the hover. Part 1. methodology and analysis. {\it J. Aerosp. Eng.} 220:61-83\\
\noindent Bandyopadhyay PR, 2009. Swimming and Flying in Nature- The Route toward Applications. {\it J. Fluids \\
\indent Eng.-Trans. ASME} 131(3):031801\\
\noindent Bao L, Hu JS, Yu YL, Peng C. et al. 2006. Viscoelastic constitutive model related to deformation of \\ 
\indent insect wing under loading in flapping motion. {\it Appl. Math. Mech.} 27: 741-8\\
\noindent Bos FM, Lentink D, Van Oudheusden BW, Bijl H. 2008. Influence of kinematics on aerodynamic\\ 
\indent performance in hovering insect flight. {\it J. Fluid Mech.} 594: 341-68  \\
\noindent Brett JR, 1963. The energy required for swimming by young sockeye salmon with a comparison of\\
\indent the drag force on a dead fish.  {\it Trans. Roy. Soc. Can.} 1-IV: 441-57\\
\noindent Brett JR, 1965a. The relation of size to the rate of oxygen consumption and sustained swimming \\
\indent speeds of sockeye salmon.  {\it J. Fish. Res. Bd. Can.} 22: 1491-501 \\
\noindent Brett JR, 1965b. The swimming energetics of salmon. {\it Sci. Am.} 213: 80-5 \\
\noindent Cheng JY, 1998. A continuous dynamic beam model for swimming fish. {\it Philos. Trans. R. Soc. Lond. B}\\
\indent 353: 981-97\\
\noindent Chopra MG, 1976. Large amplitude lunate-tail theory of fish locomotion.{\it J. Fluid Mech.} 74:161-82\\
\noindent Chopra MG, \& Kambe T, 1977. Hydrodynamics of lunate-tail swimming propulsion. Part 2. \\
\indent {\it J. Fluid Mech.} {\bf 79}, 49-69\\
\noindent Dickinson MH, G$\ddot{o}$tz KG, 1993. Unsteady aerodynamic performance of model wings at low Reynolds\\
\indent number. {\it J. Exp. Biol.} 174:45-64\\
\noindent Dickinson MH, Lehmann FO, Sane SP, 1999. Wing rotation and the aerodynamic basis of insect flight.\\
\indent {\it Science} 284: 1954-60 \\
\noindent Ellington CP, van den Berg C. Willmott AP, Thomas ALR 1996. Leading edge vortices in insect flight.\\
\indent {\it Nature.} 384: 626-30 \\
\noindent Garrick IE, 1936. Propulsion of a flapping and oscillating airfoil. NACA Rep. TR 567\\
\noindent Gray J, 1936. Studies in animal locomotion. {\it J. Exp. Biol.}. 13: 192-99 \\
\noindent Gray J, 1968. {\it Animal Locomotion.} London: Weidenfeld \& Nicolson\\
\noindent Hamdani H, \& Sun M, 2000. Aerodynamic forces and flow structures of an airfoil in some unsteady \\
\indent  motions at small Reynolds number. {\it Acta Mech.} 145: 173-87\\
\noindent Hou TY, Stredie VG, Wu TY, 2006.~A 3D numerical method for studying vortex formation behind \\
\indent  a moving plate. {\it Commun. Comput. Phys.} 1:207-28 \\
\noindent Hou TY, Stredie VG, Wu TY, 2007.~Mathematical modeling and simulation of aquatic and aerial \\
\indent  animal locomotion. {\it J. Comput. Phys.} 225: 1603-31 \\
\noindent Hu WR, Yu YL, Tong BG, Liu H, 2004.  A numerical and analytical study on a tail-flapping model\\
\indent for fish fast C-start.  {\it Acta Mech. Sini.} 20: 16-23\\
\noindent K\'{a}rm\'{a}n Th. von \& Burgers JM, 1934. General aerodynamic theory: Perfect fluids. In {\it Aerodynamic \\
\indent Theory}, Vol. II, Div. E (see Chap. V, pp.280-310)(W. F. Durand, ed.), Berlin.: Springer-Verlag\\
\noindent K\'{a}rm\'{a}n Th. von \& Sears WR, 1938. ~Airfoil theory for non-uniform motion. {\it J. Aero. Sci.}~5:379-90\\
\noindent K\'{a}rm\'{a}n Th. von \& Gabrielli G, 1950. What price speed? Specific power required for propulsion of\\
\indent vehicles. {\it Mech. Eng.} 72:775-81\\
\noindent Kooschesfahani MM, 1989. Vortical patterns in the wake of an oscillating airfoil. {\it AIAA J,}~27:1200-05\\
\noindent K$\ddot{u}$ssner, HG, 1935. Augenblicklicher Entwicklungs-stand der Frage des Fl$\ddot{u}$gelflatterns, \\
\indent {\it Luftfahrtforschung,} 6:193-209\\
\noindent Kuethe AM, 1975. On the mechanics of flight of small insects. See Wu et al. 1975, pp.803--14\\
\noindent Lai JCS, Yue J, Platzer LF, 2002. Control of backward-facing step flow using a flapping foil. {\it Exp.\\
\indent Fluids} 32:44-54\\
\noindent Lang TG, Daybell DA, 1963. {\it NAVWEPS Report} 8060, NOTS Tech. Publ. 3063\\
\noindent Levy DE, \& Seifert A, 2009. Simplified dragonfly airfoil aerodynamics at Reynolds numbers below \\
\indent 8000. {\it Physics of Fluids} 21:071901\\
\noindent Lian YS, Shyy W, Viiru D. Zhang B 2003. Membrane wing aerodynamics for micro air vehicle.  {\it Prog.\\
\indent  Aerosp. Sci.} 39: 425-65\\
\noindent Lighthill MJ, 1960. Note on the swimming of slender fish.  {\it J. Fluid Mech.} 9:305-17\\
\noindent Lighthill MJ, 1969. Hydrodynamics of aquatic animal propulsion.  {\it Ann. Rev. Fluid Mech.} 1: 413-46\\
\noindent Lighthill MJ, 1970. Aquatic animal propulsion of high hydromechanical efficiency. {\it J. Fluid Mech.}\\
\indent 44:265-301\\
\noindent Lighthill MJ, 1971. Large-amplitude elongated-body theory of fish locomotion.  {\it Proc. R. Soc. Lond.}\\
\indent B179:125-38\\
\noindent Lighthill MJ, 1973a. Aquatic animal locomotion. {\it Theoretical and Applied Mechanics} (Ed. E. Becker, \\
\indent \& G.K. Mikhailov) pp. 29-46  ~Berlin: Springer \\
\noindent Lighthill MJ, 1973b. On the Wies-Fogh mechanism of lift generation. {\it J. Fluid Mech.} 60: 1-17 \\
\noindent Lighthill MJ, 1974. Aerodynamic aspects of animal flight. In {\it Fifth Fluid Science Lecture}, pp.423-91.\\
\indent Bedford: Br. Hydrodyn. Res. Assoc. \\
\noindent Lighthill MJ, 1975. {\it Mathematical Biofluiddynamics.} Philadelphia:~SIAM\\
\noindent Lighthill MJ, 1976. Introduction to the scaling of aerial locomotion.  In {\it Scale Effects in Animal \\
\indent Locomotion} (ed. T.J. Pedley). pp.365-404 London: Academic \\
\noindent Lighthill MJ, \& Blake, R. (1990a) Biofluiddynamics of balistiform and gymnotiform locomotion. Part\\
\indent 1. Biological background, and analysis by elongated-body theory.  {\it J. Fluid Mech.} 212:183-207\\
\noindent Lighthill MJ, \& Blake, R. (1990b) Biofluiddynamics of balistiform and gymnotiform locomotion. \\
\indent Part 2. The pressure distribution arising in two-dimensional irrotational flow from general \\
\indent symmetrical motion of a flexible flat plate normal to itself.  {\it J. Fluid Mech.} 213:1-10\\
\noindent Liu H, Ellington CP, Kawachi K. van den Berg C. Willmott AP, 1998. A computational fluid dynamics\\
\indent study of hawkmoth hovering. {\it J. Exp. Biol.} 201: 461-77\\
\noindent Liu H, 2009. Integrated modeling of insect flight. {\it J. Comp. Phys.} 228: 439-59\\
\noindent Lu Y, Shen GX, \& Su WH 2007. Flow visualization of dragonfly hovering via an electromechanical\\
\indent  model. {\it AIAA Journal} 45: 615-22\\
\noindent Luo GY, \& Sun M, 2005. The effects of corrugation and wing planeform on the aerodynamic force \\
\indent  production of sweeping model insect wings. {\it Acta Mech. Sinica} 21:531-41\\
\noindent Maxworthy T, 1981. The fluid dynamics of insect flight. {\it Ann. Rev. Fluid Mech.} 13: 329-50 \\
\noindent Maxworthy T, 2007. The formation and maintenance of a leading-edge vortex during the forward\\
\indent motion of an animal wing. {\it J. Fluid Mech.} 587:471-75\\
\noindent Miloh T, 1974. The ultimate image singularities for external ellipsoidal harmonics. {\it SIAM J. Appl.\\
\indent   Math.} 26: 334-44\\
\noindent Newman JN, 1973.  The force on a slender fish-like body. {\it J. Fluid Mech.} 58:689-702\\
\noindent Newman JN, \& Wu TY, 1973. A generalized slender-body theory for fish-like forms. {\it J. Fluid Mech.}\\
\indent  57:673-93\\
\noindent Newman JN, \& Wu TY, 1975. Hydromechanic aspects of fish swimming. See Wu et al 1975, 729--62\\
\noindent Osborne, FMF, 1960. The hydrodynamical performance of migratory salmon. {\it J. Exp. Biol.} 38: 365-90\\
\noindent Pedley TJ, \& Hill SJ, 1999. Large-amplitude undulatory fish swimming: Fluid mechanics coupled to \\
\indent internal mechanics. {\it J. Exp. Biol.} 202: 3431-8\\
\noindent Pullin, DI, \& Wang JZ, 2004. Unsteady forces on an accelerating plate and application to hovering \\
\indent insect flight. {\it J. Fluid Mech.} 509: 1-21\\
\noindent Pyatetskiy VY, 1970. Hydrodynamic Problems.  {\it Bionics} 4: 12-23, Transl. from Russian, 1971,  \\
\indent Joint Pub. Res. Service 52605, NTIS.\\
\noindent Ringuettet MJ, Milano M, Gharib M. 2007. Role of the tip vortex in the force generation of low-aspect-\\ 
\indent ratio normal flat plates. {\it J. Fluid Mech.} 581: 453-68\\
\noindent Schmidt-Nielsen K, 1972. Locomotion: energy cost of swimming, flying, running. {\it Science} 177:222-8\\
\noindent Singh K, \& Pedley TJ, 2008. The hydrodynamics of flexible-body manoeuvres in swimming fish. \\
\indent {\it Physica D-Nonlinear Phenomena} 237: 2234-9\\
\noindent Sohn MH, \& Chang JW, 2007. Flow visualization and aerodynamic load calculation of three types\\
\indent of clap-fling motions in a Weis-Fogh mechanism. {\it Aerosp. Sci. Technol.} 11: 119-29\\
\noindent Stredie VG, 2005. ~Mathematical modeling and simulation of aquatic and aerial animal locomotion.\\
\indent ~Ph.D. Thesis, California Institute of Technology, Pasadena, CA \\
\noindent Su Y, Wu TY, \& Yates GT, 1983. The effect of sidefin vortex sheet on fish propulsion. (unpublished)\\
\indent Northwestern Polytechnic University, Xian, Shaanxi, China.\\
\noindent Sun M, \& Tang J, 2002. Unsteady aerodynamic force generation by a model fruit fly wing in flapping \\
\indent motion. {\it J. Exp. Biol.} 205: 55-70\\
\noindent Taira K, \& Colonius T, 2009. Three-dimensional flow around low aspect-ratio flat plate wings at low \\
\indent  Reynolds numbers. {\it J. Fluid Mech.} 623: 187-207\\
\noindent Taylor GI, 1952. Analysis of the swimming of long and narrow animals. {\it Proc. R. Soc. A} 214:158-83 \\
\noindent Theodorsen T, 1935. General theory of aerodynamic instability and mechanism of flutter. NACA \\
\indent Tech. Report 496\\
\noindent Tong BG, Lu XY, 2004. A review on biomechanics of animal flight and swimming. {\it Advances in \\
\indent  Mechanics}, Chinese Academy of Sciences. 34(1):1-8\\
\noindent Triantafyllou MS, Triantafyllou GS, Yue DKP, 2000. Hydrodynamics of fishlike swimming. {\it Ann. Rev. \\
\indent Fluid Mech.} 32: 33-53 \\
\noindent Tuck EO, 1975. The effect of span-wise variations in amplitude on the thrust-generating performance \\
\indent of a flapping thin wing. See Wu et al. 1975, pp.953--74\\ 
\noindent Tucker VA, 1975. The energy cost of moving about.  {\it Am. Scientist} 63: 413-9\\
\noindent Wagner H, 1925. ~$\ddot{U}$ber die Entstehung des dynamischer Austrieb von Tragfl$\ddot{u}$geln. {\it ZAMM} 5:17-35\\
\noindent Wang ZJ, 2005. Dissecting insect flight. {\it Annu. Rev. Fluid Mech.} 37: 183-210 \\
\noindent Webb PW, 1975. Hydrodynamics and energetics of fish propulsion. {\it Bull. Fish. Res. Bd. Can.}190: \\
\indent 190: 1-158.\\
\noindent Weihs D, 1972. A hydrodynamic analysis of fish turning maneuvres. {\it Proc.R.Soc.Lond.}B182:59-72.\\
\noindent Weihs D, 1973. The mechanism of rapid starting of slender fish.  {\it Biorheology} 10:343-50\\
\noindent Weinblum G, 1954. Approximate theory of heaving and pitching of hydrofoils in regular shallow waves. \\
\indent DTMB Report C-479.  David Taylor Model Basin, Washington, DC \\
\noindent Weis-Fogh T, 1973. Quick estimates of flight fitness in hovering animals, including novel mechanism \\
\indent for lift production. {\it J. Exp. Biol.} 59: 169-230\\
\noindent Weis-Fogh T, 1975. Hovering flight of the dragonfly {Aeschna Juncea L.} kinematics and aerodynamics\\
\indent See Wu, Brokaw et al 1975, pp.729-62 \\
\indent New York: Plenum Press\\
\noindent Weis-Fogh T, \& Jensen M, 1956. Biology and physics of locust flight, I-IV. {\it Philos. Trans. Roy Soc.\\
\indent  Lond.} B239: 415-584\\
\noindent Wu DM, Sheng QH, Zhang L, 2005.~Aerodynamic forces acting on an albatross flying above sea waves.\\  
\indent {\it Appl. Math. \& Mech.} 26:1114-20\\
\noindent Wu TY, 1961.~Swimming of a waving plate. {\it J. Fluid. Mech.} 10:321-44\\
\noindent Wu TY, 1962.~Accelerated swimming of a waving plate. {\it Proc. of 4th Symp. on Naval Hydrodynamics}, \\
\indent pp.457-473, Publ.ONR.ACR-92, Washington DC: Academy \\
\noindent Wu TY, 1966.~The mechanics of swimming.  In {\it Biomechanics, Proc. of Symp. on Biomech.} (ed. Y. C. \\
\indent Fung) pp. 187-204. New York: ASME \\
\noindent Wu TY, 1968.~Fluid mechanics of swimming propulsion.  {\it Proc. of 7th Symp. on Naval Hydrodynamics},\\
\indent (eds. R.D. Cooper, S.W. Doroff) 1173-201, ONR.DR-148, Washington DC: Academy \\
\noindent Wu TY, 1971a. Hydromechanics of swimming propulsion.  Part 1: Swimming of a two-dimensional\\
\indent flexible plate at variable forward speeds in an inviscid fluid. {\it J. Fluid Mech.} 46: 337-55\\
\noindent Wu TY, 1971b. Hydromechanics of swimming propulsion.  Part 2: Some optimum shape problems. \\
\indent {\it J. Fluid Mech.} 46: 521-44\\
\noindent Wu TY, 1971c. Hydromechanics of swimming propulsion.  Part 3:  Swimming and optimum movements \\
\indent  of slender fish with side fins.  {\it J. Fluid Mech.} 46:545-68\\
\noindent Wu TY, 1971d. Hydrodynamics of swimming fishes and cetaceans.  In {\it Advances in Applied Mechanics}, \\
\indent (ed. C. S. Yih) 11:1-63.  New York: Academic \\
\noindent Wu TY, 1972.  Extraction of flow energy by a wing oscillating in waves. {\it J. Ship Research}, 14: 66-78\\
\noindent Wu TY, 1976. Introduction to the scaling of aquatic animal locomotion.  In {\it Scale Effects in Animal \\
\indent Locomotion} (ed. T.J. Pedley). 203-232. London: Academic \\
\noindent Wu TY, 1980. Extraction of energy from wind and ocean current. {\it Proc. of 13th Symp. on Naval \\
\indent Hydrodynamics} (eds. S. Schuster, M. Schmiechen) 2:1-10, ONR.ACR-92, National Academy \\
\noindent Wu TY, 1984. Mechanophysiology of aquatic animal swimming.  In {\it Biomechanics in China, Japan, and \\
\indent U.S.A.} (eds. Y.C. Fung, E. Fukada, Wang Junjian). pp. 291-304. Beijing: Science  \\
\noindent Wu TY, 2001a. On theoretical modeling of aquatic and aerial animal locomotion. In {\it Adv. Appl.\\
\indent Mech.} 38:291-353, Boston: Academic \\
\noindent Wu TY, 2001b. Mathematical biofluiddynamics and mechanophysiology of fish locomotion. (3rd Ann.  \\
\indent Meeting, Israeli Soc. Theoretical and Mathematical Biology, 5 Jan. 2000.) In {\it Biofluiddynamics --\\
\indent In memory of Sir James Lighthill,} a special issue of {\it J. Math. Meth. Appl. Sci.}  24: 1541-64\\
\noindent Wu TY, 2005.~Reflections for resolution to some recent studies on fluid mechanics. In {\it Advances in \\
\indent ~Engineering Mechanics -- Reflections and Outlooks.}~(ed. AT Chwang, MH Teng, DT Valentine)\\
\indent  pp.693-714 ~ Singapore: World Sci. \\
\noindent Wu TY, 2006.~A nonlinear unsteady flexible wing theory.  {\it Struct. Control Health Monit.} 13: 553-60 \\
\noindent Wu TY, 2007. ~A nonlinear theory for a flexible unsteady wing.  {\it J. Eng. Math.} 58:279-87 \\
\noindent Wu T.Y. Wu, Brokaw CJ, Brennen C. eds. 1975 ~{\it Swimming and Flying in Nature,} Vol. 2. New York: Plenum \\
\noindent Wu TY, \& Chwang AT, 1974. Double-body flow theory - A new look at the classical problem. {\it Proc. \\
\indent the 10th Symp. on Naval Hydrodynamics} (ed. R. D. Cooper) pp. 89-106. Washington DC: Academy\\
\noindent Wu TY, \& Chwang AT, 1975. Extraction of flow energy by fish and birds in a wavy stream. \\
\indent See Wu et al. 1975, pp. 687-702. \\
\noindent Wu TY, \& Newman, JN, 1972. Unsteady flow around a slender fish-like body. {\it J. Mech. Eng. Sci.} \\
\indent  14(7 Suppl.): pp.43-52 \\
\noindent Wu TY, \& Yates GT, 1978. A comparative mechanophsiological study of fish locomotion with \\
\indent  implication for tuna-like swimming mode.  In {\it The Physiological Ecology of Tuna} (eds. GD Sharp,\\
\indent  AE Dizon), pp. 313-38.  New York: Academic \\
\noindent Yang Y, Wu GH, Yu YL, Tong BG, 2008. Two-dimensional self-propelled fish motion: an integrated \\
\indent method for deforming body dynamics and unsteady fluid dynamics. {Chin. Phys. Let.} 25:560-97\\
\noindent Yates GT, 1977. {\it Finite amplitude unsteady slender body theory and experiments.}  Ph.D. Thesis, \\
\indent Calif. Inst. Technol. Pasadena, CA \\
\noindent Yates GT, 1983. Hydromechanics of body and caudal fin propulsion. Chapter 6 in {\it Fish Biomechanics} \\
\indent (eds. PW Webb \& D Weihs), pp.177-213. New York: Praeger Sci. \\
\noindent Yu Y, Tong B, \& Ma H, 2003. An analytic approach to Theoretical modeling of highly unsteady viscous\\
\indent flow excited by wing flapping in small insects.  {\it ACTA Mech. Sinica} 19(6): 508-16\\
\noindent Yu Y, Tong B, 2005. A flow control mechanism in wing flapping with stroke asymmetry during insect\\
\indent forward flight. {\it ACTA Mech. Sinica} 21(3): 218-27\\
\noindent Yu Y, Tong B, \& Ma H, 2005. Unsteady flow mechanics revisited in insect flapping flight. {\it  ACTA\\ 
\indent Mech. Sinica.} 37(3): 257-65\\
\end{document}